\documentclass[iop,revtex4]{emulateapj}
\usepackage{amsmath}
\usepackage{color}
\usepackage{lscape}

\newcommand{\Msolar}{M$_{\odot}$}
\newcommand{\kms}{km~s$^{-1}$}
\newcommand{\eoi}{$e/i$}
\newcommand{\PRV}{$P_{\rm RV}$}
\newcommand{\PPM}{$P_{\rm PM}$}

\newcommand{\NRV}{13776}   %Number of RVs total (NRVC+NRVW)
\newcommand{\NRVC}{8023}   %Number of CfA RVs
\newcommand{\NRVW}{5753}   %Number of WIYN RVs
\newcommand{\Nstars}{1278} %Number of stars observed
\newcommand{\NstarsC}{455} %Number of stars observed at CfA
\newcommand{\NstarsW}{1163} %Number of stars observed at WIYN
\newcommand{\NMEM}{562}    %Number of cluster members (SM+BM+BLM+BU)
\newcommand{\NVAR}{142}    %Number of BM+BLM+BU
\newcommand{\SM}{420}      %Number of SM
\newcommand{\SN}{498}      %Number of SN
\newcommand{\BM}{108}       %Number of BM
\newcommand{\BN}{17}       %Number of BN
\newcommand{\BLM}{23}      %Number of BLM
\newcommand{\BLN}{58}      %Number of BLN
\newcommand{\BU}{11}       %Number of BU
\newcommand{\U}{143}       %Number of U

\begin{document}

\title{Stellar Radial Velocities in the Old Open Cluster M67 (NGC 2682) I. Memberships, Binaries, and Kinematics \footnote{WIYN Open Cluster Study. LXVII.}}
\shorttitle{Stellar Radial Velocities in the Old Open Cluster M67}

\author{Aaron M. Geller\footnote{Visiting Astronomer, Kitt Peak National Observatory, National Optical Astronomy Observatory, which is operated by the Association of Universities for Research in Astronomy (AURA) under cooperative agreement with the National Science Foundation.}$^,$\footnote{NSF Astronomy and Astrophysics Postdoctoral Fellow}}
\affil{Center for Interdisciplinary Exploration and Research in Astrophysics (CIERA) and Department of Physics \& Astronomy, Northwestern University, Evanston, IL 60201, USA, \\ Department of Astronomy and Astrophysics, University of Chicago, 5640 S. Ellis Avenue, Chicago, IL 60637, USA; a-geller@northwestern.edu}

\author{David W. Latham}
\affil{Harvard-Smithsonian Center for Astrophysics, 60 Garden Street, Cambridge, MA 02138, USA; dlatham@cfa.harvard.edu}

\author{~Robert D. Mathieu$^\dagger$}
\affil{Department of Astronomy, University of Wisconsin - Madison, WI 53706, USA; mathieu@astro.wisc.edu}

\shortauthors{Geller, Latham \& Mathieu}

\begin{abstract}  

We present results from \NRV\ radial-velocity measurements of \Nstars\ candidate members
of the old (4 Gyr) open cluster M67 (NGC 2682). 
The measurements are the results of a long-term survey that includes data from 
seven telescopes with observations for some stars spanning over 40 years.
For narrow-lined stars, radial velocities are measured with precisions
ranging from about 0.1 to 0.8~\kms. 
The combined stellar sample reaches from the brightest giants in the cluster down to about 4 magnitudes below the main-sequence 
turnoff ($V = 16.5$), covering a mass range of about 1.34~\Msolar\ to 0.76~\Msolar.
Spatially, the sample extends to a radus of 30 arcminutes
(7.4 pc in projection at a distant of 850 pc or 6-7 core radii). 
We find M67 to have a mean radial velocity of $+33.64$~\kms\ (with an internal precision of $\pm 0.03$~\kms)
well separated from the mean velocity of the field.  For  
stars with $\geq$3 measurements, we derive radial-velocity membership 
probabilities and identify radial-velocity variables, finding \NMEM\ cluster 
members, \NVAR\ of which show significant radial-velocity variability.  
We use these cluster members to construct a color-magnitude diagram and 
identify a rich sample of stars that lie far from the standard single 
star isochrone, including the well-known blue stragglers, sub-subgiants and yellow giants. 
These exotic stars have a binary frequency of (at least) 80\%, more than three times that detected for stars in the remainder of the sample.
We confirm that the cluster is mass segregated, finding the 
binaries to be more centrally concentrated than the single stars in our sample at the 99.8\% confidence level 
(and at the 98.7\% confidence level when only considering main-sequence stars).
The blue stragglers are centrally concentrated as compared to the solar-type main-sequence single stars in the 
cluster at the 99.7\% confidence level.
Accounting for measurement precision, we derive a 
radial-velocity dispersion in M67 of $0.80 \pm 0.04$~\kms\ for our sample of single main-sequence stars, subgiants and giants with $V \leq 15.5$.
When corrected for undetected binaries, this sample yields a true radial-velocity dispersion of $0.59 \substack{+0.07\\-0.06}$~\kms.
The radial distribution of the velocity dispersion
is consistent with an isothermal distribution within our stellar sample.
Using the cluster radial-velocity dispersion, we estimate a virial mass for the cluster of 
$2100 \substack{+610\\-550}$~\Msolar.

\end{abstract}

\keywords{open clusters and associations: individual (NGC 2682) -- binaries: spectroscopic -- methods: observational -- techniques: spectroscopic -- blue stragglers -- stars: kinematics and dynamics}

\section{Introduction}
\label{sec:intro}

M67 (NGC~2682) is one of the few old and rich open clusters in our Galaxy, and therefore is central
to our understanding of stellar evolution, stellar dynamics and star cluster evolution.
M67 is located at $\alpha = 8^h 51^m 23\fs3, \delta = +11\degr 49 \arcmin 02\arcsec$ (J2000).
With an age of about 4 Gyr \citep{nissen:87,montgomery:93,demarque:92,carraro:94,fan:96,vandenberg:04,balaguer:07} and 
a half-mass relaxation time of about 100 Myr \citep{mathieu:86b}, M67 is a highly dynamically evolved system.  
Importantly, M67 is relatively nearby, with recent distance measurements ranging from between about 800 pc to 900 pc 
\citep{janes:85,nissen:87,montgomery:93, carraro:94,fan:96,grocholski:03,sandquist:04,balaguer:07,pasquini:08,sarajedini:09}, 
and has low extinction, with recent $E(B-V)$ determinations between 0.015 and 0.056
\citep{janes:84,burstein:86,montgomery:93,carraro:94,fan:96,taylor:07}.
Furthermore, most studies agree that M67 has a roughly solar metallicity, with recent [Fe/H] values ranging from 
-0.10 to +0.05 \citep{janes:84,burstein:86,nissen:87,montgomery:93,fan:96,randich:06,balaguer:07,taylor:07,friel:10,pancino:10,jacobson:11}.
Thus M67 provides a large sample of solar-type dwarf and evolved stars that are easily accessible to a variety of ground-based 
(and space-based) observations.

Indeed, M67 has been extensively studied through photometry from X-rays to near-IR
\citep{belloni:98,vandenberg:04,landsman:98,nissen:87,montgomery:93,fan:96,balaguer:07,yadav:08,sarajedini:09}, including several time-series 
optical photometric surveys \citep{gilliland:91,gilliland:93,vandenberg:02,stassun:02,sandquist:03b,bruntt:07,pribulla:08,yakut:09}.
These surveys have revealed many intriguing stars, some of which lie far from the standard single-star evolutionary sequence in a color-magnitude 
diagram (CMD).  

In order to interpret these observations, kinematic membership probabilities are paramount.  There have been a number of proper-motion surveys 
of M67 (\citealt{sanders:77,girard:89,zhao:93,yadav:08}, and see also \citealt{loktin:05}), to various limiting magnitudes and 
spatial extents.  There have also been a few radial-velocity (RV) surveys \citep{mathieu:86,mathieu:90,milone:92,milone:94,yadav:08,pasquini:11}.
Surveys such as these have confirmed the kinematic cluster memberships of a rich population of blue straggler stars (BSS; residing blueward of and generally brighter than the 
main-sequence turnoff), yellow giants (residing between the BSS region and the normal giant sequence), and 
``sub-subgiants'' (residing to the red of the main sequence but fainter than the subgiant branch, and also known as ``red stragglers''),
among the kinematic cluster members (see Figure~\ref{fig:cmd} and Section~\ref{sub:snote}).

Numerous theoretical efforts have aimed to explain the origins of these exotic stars through studies of the dynamical evolution of M67
\citep[e.g.][]{leonard:92,leonard:96,hurley:05}. These studies show that close stellar encounters, and particularly those involving 
binary stars, may be relatively frequent in M67 and can lead to the creation of exotic stars similar to those 
observed in the true cluster.  Furthermore, these models emphasize the importance of binaries to the dynamical evolution of the 
cluster.

To date, the published results of the binary population in M67 has been limited.  The largest study of binaries in M67 is that of 
\citet{mathieu:90}, who observed a sample of bright ($V < 12.8$) proper-motion members, and present orbits for 22 spectroscopic binaries.  
We have continued to monitor these and other stars in M67  in order to extend our sample of detected binaries (and 
those with orbital solutions) to longer orbital periods, fainter magnitudes and a larger distance from the cluster center. 
A progress report summarizing the characteristics of 85 spectroscopic binaries was presented at the General Assembly of the 
IAU in Prague \citep{latham:07}.

Here we present results from our ongoing RV survey of the cluster.  To date we have obtained \NRV\ RV measurements\footnote{Throughout this paper, when quoting numbers of RV measurements, we provide the number of RVs from the primary stars (and therefore count the number of spectra that result in at least one RV measurement).  We do not add to our count additional RVs from, for instance, the secondaries of SB2s.} of 
\Nstars\ stars in M67 with $8 \leq V \leq 16.5$ (about 1.34~\Msolar\ to 0.76~\Msolar) and extending spatially to 30 arcmin in radius from the cluster center
(7.4 pc in projection at a distance of 850 pc, or approximately 6 to 7 core radii).\footnote{\citet{zhao:96} derive a core radius for M67 of 5.2 arcmin.
\citet{bonatto:03} use 2MASS data to derive a core radius for M67 of 4.86 arcmin.
\citet{davenport:10} revised this result by reanalyzing 2MASS data, and find a core radius of 4.12 arcmin (but a much larger 
core radius of $8.24 \pm 0.60$ arcmin when using the fainter stars in their SDSS sample).  Therefore our sample extends to
between approximately 6 and 7 core radii.} 
Our stellar sample spans from the brightest stars in the cluster down to 4 magnitudes below the main-sequence turnoff.
Also for reference, \citet{fan:96} find a half-mass radius for stars in our observed magnitude range in M67 of about 10 to 11 arcmin.
Tidal radius estimates for the cluster range from 50 arcmin to 100 arcmin \citep{keenan:73,piskunov:08,davenport:10}.

In Section~\ref{sec:sample}, we define our stellar sample in detail, and in Sections~\ref{sec:obs}~and~\ref{sec:data} we describe our 
observations and the completeness of our data.
In short, our observations are nearly complete within our ``primary sample'' of stars with $V \leq 15.5$ and within 30 arcmin from the cluster center.  
Within this primary sample, 
we have at least one RV measurement for all but one star and at least 3 RV measurements for all but four stars (two of which are rapid rotators).
None of these four stars with $<3$ RV measurements in our primary sample are proper-motion members;
moreover, we have at least three RV measurements for all proper-motion members in our primary stellar sample.  
Our time baseline of observations for some stars reaches to over 40 years.

Throughout this paper, we use a cutoff of at least 3 RV measurements before attempting to derive RV membership probabilities and 
variability statistics.  Extensive Monte Carlo analyses by \citet{mathieu:83} and \citet{geller:12} indicate that 3 RV measurements over the course of 
at least one year is sufficient to detect nearly all binaries with orbital periods of less than $10^3$ days.  With our observations, we can detect binaries 
with significantly longer periods, out to $\sim10^4$ days.  The detection of binaries is 
particularly important for determining reliable RV membership probabilities.  In general, binaries without RV orbital solutions
do not yield precise RV membership probabilities, due to their unknown center-of-mass ($\gamma$-) RVs.  We discuss our identification of 
binaries and cluster members in Section~\ref{sec:results}, and we present a summary table of our results for each observed star in Table~\ref{RVtab}.

Finally, in Section~\ref{sec:disc} we use our cluster members to investigate the CMD (cleaned from 
field star contamination), a few stars of note, the projected radial distribution of different cluster populations, and the velocity 
dispersion and virial mass of the cluster.  Then in Section~\ref{sec:summary} we summarize our results.

\section{Stellar Sample}
\label{sec:sample}

\begin{table*}[!t]
\centering
\begin{minipage}{0.45\textwidth} %silly fix, but at least this works
\caption{\label{t:tel}Time Span of Data from Each Telescope \vspace{0.2em}}
\end{minipage}
\begin{tabular}{lccccc}
\hline
\hline
\relax\\[-1.7ex]
Telescope & HJD$_{\rm start}$ & HJD$_{\rm end}$ & Days & $N_{\rm stars}$ & $N_{\rm obs}$ \\
\relax\\[-1.7ex]
\hline
\relax\\[-1.7ex]
Palomar Hale 5m          & 2440952 & 2445297 &  4345 &  112 &  367 \\
Tillinghast 1.5m + DS    & 2444184 & 2454958 & 10774 &  357 & 4670 \\
CORAVEL                  & 2444340 & 2446413 &  2073 &    5 &   92 \\
MMT + DS                 & 2445337 & 2450830 &  5493 &  376 & 1889 \\
Wyeth 1.5m + DS          & 2445722 & 2453433 &  7711 &   34 &  332 \\
WIYN + Hydra             & 2453386 & 2456709 &  3323 & 1163 & 5753 \\
Tillinghast 1.5m + TRES  & 2455143 & 2456801 &  1658 &  111 &  672 \\
\relax\\[-1.7ex]
\hline
\relax\\[-1.7ex]
\end{tabular}
\end{table*}

Our RV survey of M67 began as part of the dissertation
work of \citet{mathieu:83}, taking advantage of the CfA Digital
Speedometers \citep[DS;][]{latham:85,latham:92}.  Three nearly identical instruments
were used, initially on the MMT and 1.5-m Tillinghast Reflector at the
Fred Lawrence Whipple Observatory on Mount Hopins, Arizona, and then
later on the 1.5-m Wyeth Reflector at the Oak Ridge Observatory in the
Town of Harvard, Massachusetts.
Subsequently the M67 
target samples were expanded several times. RV measurements from other programs were integrated into the database,
and our observational facilities were extended to include Hydra at the WIYN
Observatory\footnote{The WIYN Observatory is a joint facility of the University of Wisconsin-Madison, Indiana University, the National Optical Astronomy Observatory and the University of Missouri.} and TRES on the Tillinghast Reflector.
In Table~\ref{t:tel} we list the different telescopes and 
instruments used for this project, along with the dates, number of stars observed and number of observations from each telescope. 
For the record, here we briefly review the history 
of what now constitutes more than 40 years of observations, and end by defining the M67 stellar sample that comprises the foundation of this paper. 

\subsection{CfA}

\begin{figure}[!t]
%\epsscale{0.45}
\plotone{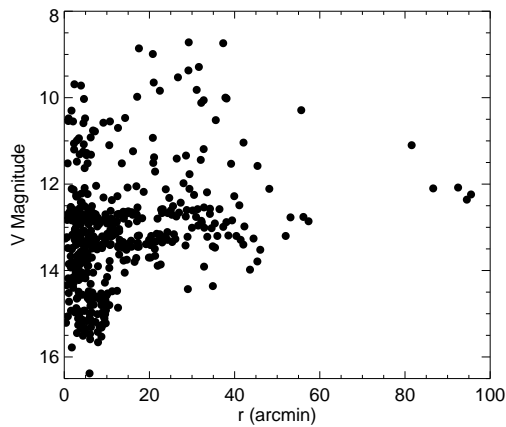} \\
\plotone{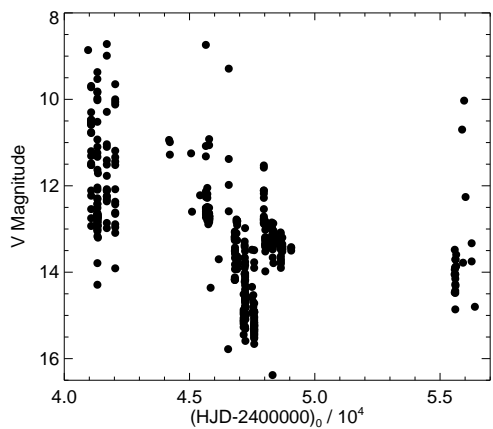} \\
\plotone{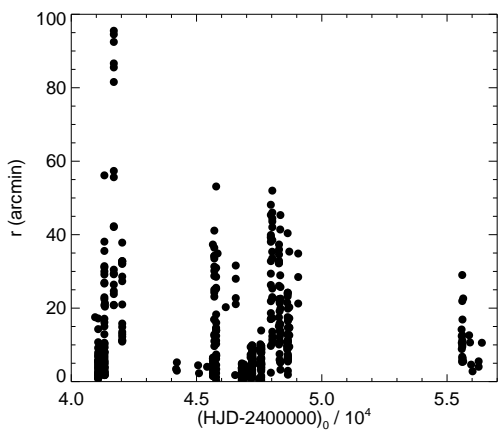} \\
\caption{\small The CfA stellar sample, showing (top) $V$ magnitude as a function of radius from the cluster center, 
and the chronology of the sample, with $V$ magnitude (middle) and radial distance from the 
cluster center (bottom) plotted as functions of the HJD of the first observation, respectively, for all CfA observed stars.}
\label{fig:CfAsamp}
\epsscale{1}
\end{figure}

After exploratory observations of a few stars, the initial CfA sample was defined in 1982 as all stars with \citet{sanders:77} proper-motion membership 
probabilities greater than 50\%, $V < 12.8$ and $(\bv) > 0.40$. This sample comprised the top of the main sequence, subgiants and the giant branch.
(The \citealt{sanders:77} study covers a square of approximately 80 arcmin on a side, centered on the cluster.)
A study to monitor the RVs of the 13 classical blue
stragglers in M67 was being pursued independently \citep{latham:96}, and the relevant results from that survey are included in this
paper as a convenience to the reader.
Also, a subset of 28 giants and subgiants were observed more intensively from Spring 1982 to Spring 1983.

By Spring 1988 the target sample expanded substantially to include all \citet{sanders:77} members to $V=14$, and all members in the cluster core 
(radius $< 10$ arcmin) to $V=16$ (only observable with the MMT), totaling 432 stars. 
The last surviving CfA Digital Speedometer, on the 1.5-m Tillinghast
Reflector, was retired in the summer of 2009.  Over the following five
observing seasons the new Tillinghast Reflector Echelle Spectrograph
(TRES) was used to continue the RV observations of targets (mostly
binaries) from both the CfA and the WIYN samples.
Whenever the observations accumulating for a target suggested that the velocity was
not constant, additional observations were scheduled at a frequency
designed to reveal its orbital parameters. For some systems the CfA
observations span almost 35 years.

Importantly, Roger Griffin and James Gunn also had a RV program for M67 from 1971 to 1982 at the Palomar Hale 5m telescope, which they supplemented with 
contemporaneous observations obtained with the CORAVEL instrument at Haute Provence for five of the binaries. Their target sample was very similar to our 
initial sample, and the combination of the two data sets to expand the time baseline was a natural and straightforward step. The integration of the Palomar 
and CORAVEL data with the CfA data is discussed in detail in \citet{mathieu:86}, where the entirety of the Palomar data as well as the CfA data to that 
date are presented, and in \citet{mathieu:90}.

Hereinafter, we will refer to the target stars and the measurements taken at all telescope/instrument pairs except WIYN/Hydra as the CfA sample and data.
As of February 2015, there are 447 stars in the CfA sample.
The radius -- $V$-magnitude distribution of the stars observed by these telescopes is shown in Figure~\ref{fig:CfAsamp} (top), and 
the middle and bottom panels of the same figure present the chronology of observations graphically.

\subsection{WIYN Stellar Sample}

\begin{figure}[!t]
%\epsscale{0.45}
\plotone{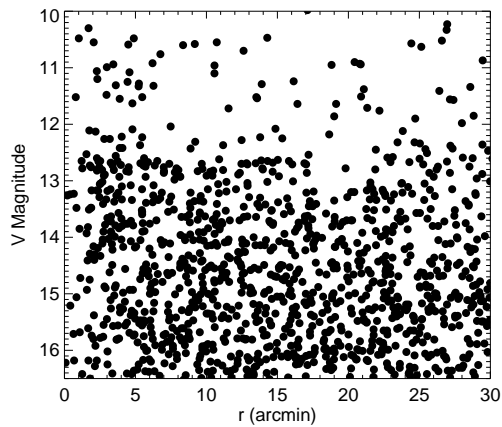} \\
\plotone{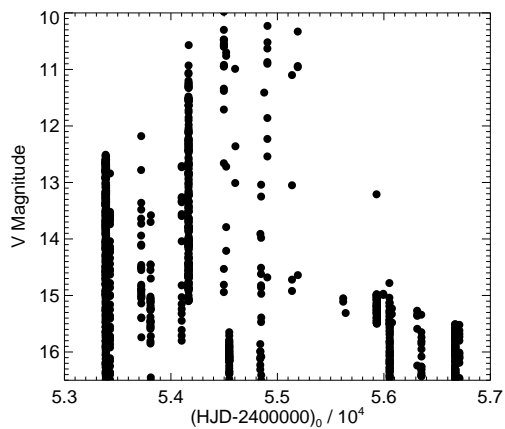} \\
\plotone{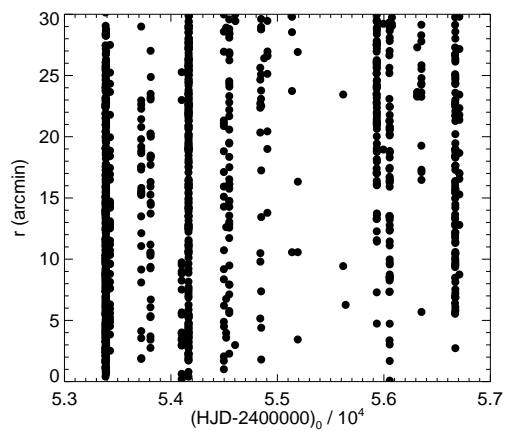} \\
\caption{\small The WIYN stellar sample, showing (top) $V$ magnitude as a function of radius from the cluster center, 
and the chronology of the sample, with $V$ magnitude (middle) and radial distance from the 
cluster center (bottom) plotted as functions of the HJD of the first observation, respectively, for all WIYN observed stars.}
\label{fig:Wsamp}
\epsscale{1}
\end{figure}

As evident in Figure~\ref{fig:CfAsamp}, the CfA sample is not comprehensive in radius for the fainter stars in the sample. Furthermore, the \citet{sanders:77} 
proper-motion study was progressively more incomplete and less precise with increasing magnitude. The Hydra Multi-Object Spectrograph (MOS) on the WIYN 3.5m telescope, 
combined with modern target lists, has been able to provide a complete magnitude and spatially limited sample, as described here.

WIYN observations of M67 began on January 15, 2005 as part of the WIYN Open Cluster Study \citep[WOCS;][]{mathieu:00}. The current WIYN target list contains all stars with $10 \leq V \leq 16.5$ and within 30 arcminutes in radius 
from the cluster center.  We drew our sample (including astrometric positions) from the 2MASS catalog, supplemented with photometry from \citet{montgomery:93}. 
The Hydra MOS on WIYN has a 1\degr\ field of view, which sets the radial extent of the WIYN survey. 

We can derive reliable RVs with our WIYN observing setup for stars with $(\bv)_0 > 0.4$. (Rapid rotation and a diminished number of absorption features 
often hinders our ability to derive precise RVs for earlier-type stars.) The only stars in M67 that are bluer than this limit and bright enough to be within our sample 
are BSS. Because of their scientific interest, we included these stars in the WIYN stellar sample. In total there are 1278 stars within the WIYN stellar sample; 
382 of these stars are also in the CfA stellar sample. We show the radius -- $V$-magnitude distribution for stars observed at WIYN in the top panel of Figure~\ref{fig:Wsamp}. 

As seen in Figure~\ref{fig:Wsamp} (middle), the $V$ magnitude range of the stars observed at WIYN has evolved with time. Initially we chose to make use of WIYN's 
advantage at faint magnitudes and prioritized stars from $12.5 < V < 16.5$. (The Hydra MOS has a dynamic range limit of 4 mag.) Later, to improve the completeness 
of our observations within the combined WIYN and CfA sample, we extended our WIYN sample to include all stars within a 30 arcminute radius from the cluster center 
and with $10 \leq V \leq 16.5$. For a few epochs, we focused exclusively on the brighter stars in our list in order to build this complete sample.

We also note that we removed from our sample the few high-proper-motion stars in the M67 field whose coordinates have changed 
significantly during the course of our survey, as these stars are certainly not cluster members.

\pagebreak
\section{Observations}
\label{sec:obs}

Details about the telescopes, observing procedures, and data
reductions of spectra obtained with the CfA Digital
Speedometers can be found in \citet{latham:85,latham:92}.  The corresponding
information for spectra obtained with Hydra at the WIYN Observatory
can be found in \citet{geller:08,geller:10} and \citet{hole:09}.  
We describe briefly here the TRES instrument and observations.

TRES is a stabilized fiber-fed echelle spectrograph with a CCD
detector and resolution of 44,000.  Procedures were adopted to ensure
that the RVs from TRES could be adjusted to the native velocity system
of the CfA Digital Speedometers.  Although TRES delivers wavelength
coverage from 390 to 900 nm, only the order centered on the Mg b
features was used to derive RVs in order to match the spectral region
used by the CfA Digital Speedometers, using the same library of
synthetic template spectra for the correlation analysis.  To establish
the zero point offset between the TRES and CfA Digital Speedometer
RVs, observations of the afternoon blue sky and of selected IAU
standards were obtained on most nights.  There were several
modifications to TRES during the first years of its operation, and
corresponding zeropoint shifts as large as 0.1~\kms\ were measured.
Since March 2012 the TRES zeropoint has been stable at the level of
about 0.01~\kms.

The TRES velocities are all (initially) shifted by the gravitational redshift of the Sun 
and blueshift of the Earth, because the library of templates does not include 
either of those effects.  Therefore the derived stellar velocities are all redshifted 
by the net amount, nominally $+0.62$~\kms, which we subtract out.  Furthermore, 
the native CfA Digital Speedometer velocity system is shifted by $-0.14$~\kms\ compared to the IAU system \citep{stefanik:99}\footnote{Note that the offset of the CfA Digital
Speedometer native velocity system reported in that paper has the wrong sign.}.  
The actual correction to get the TRES velocities onto the native CfA system is observed to be $-0.75$~\kms, 
very close to $-0.62 - 0.14 = -0.76$~\kms.  We report TRES velocities on this native CfA system.

We do not apply any shift to the WIYN RVs, and, as noted in Section~\ref{sec:data}, we do not detect any significant zero-point
shift between the WIYN RVs and the native CfA system.  Therefore by construction, all RVs reported here, as well as the mean RV of the cluster, are 
on the native CfA system.

\subsection{Radial-Velocity Precision}
\label{sub:prec}

In order to detect binaries (and higher-order systems), we require multiple observations
at multiple epochs with known RV precisions.  Our sample contains observations of mainly narrow-lined stars 
within a modest magnitude range, but taken at multiple different telescopes with different instruments.  
Therefore to facilitate our subsequent analysis of binarity (Section~\ref{sub:vvar}), we estimate
single-measurement precision values for RVs derived from each respective telescope.  

\begin{figure}[!t]
%\epsscale{0.5}
\plotone{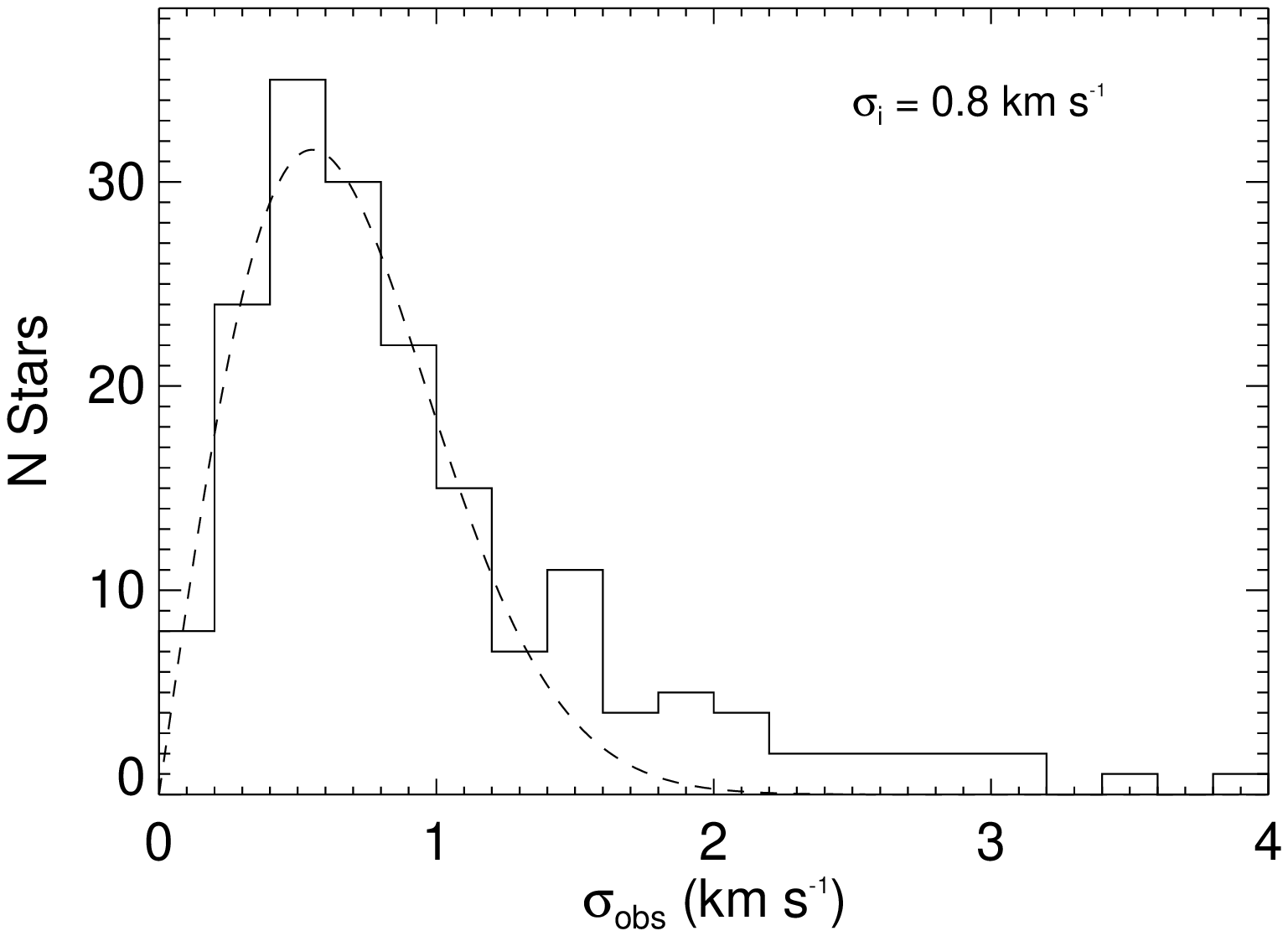}\\
\plotone{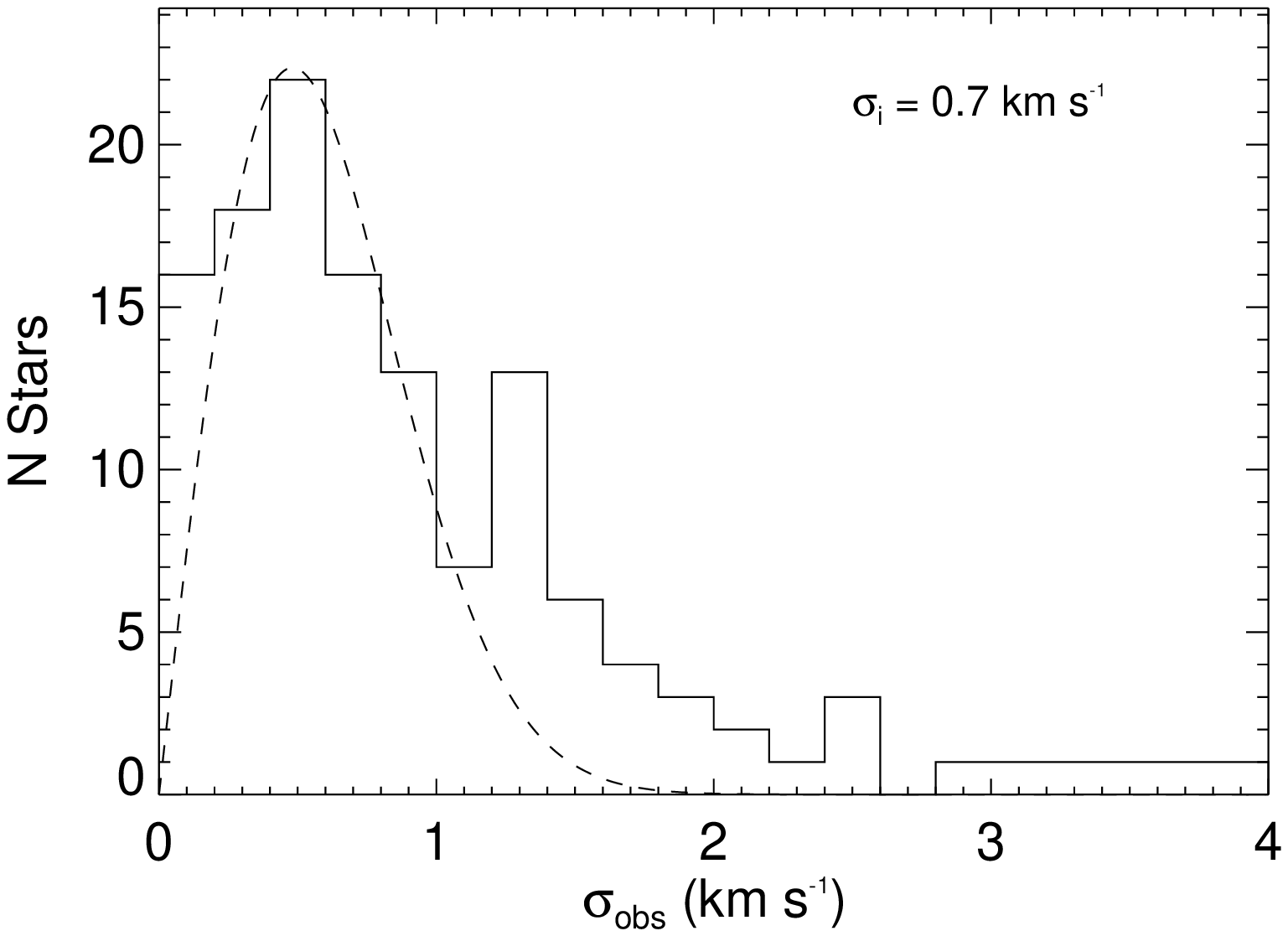}\\
\plotone{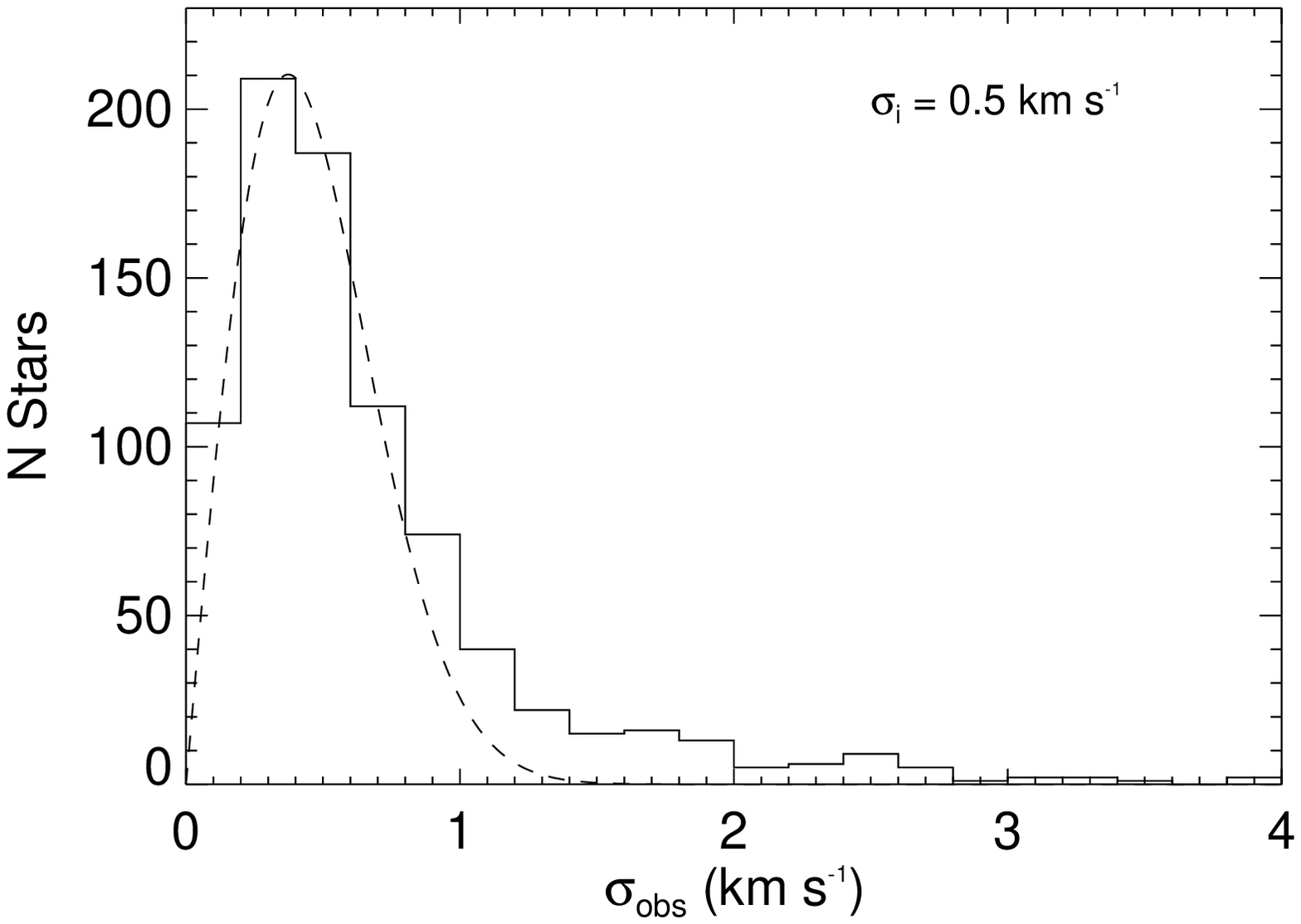}\\
\caption{\small Histograms of the RV standard deviations of the first three 
RV measurements for all stars observed $\geq3$ times at the Tillinghast Reflector + DS (top), MMT (middle) and the WIYN (bottom) telescopes. Also 
shown in the dashed line in each panel is the best fitting $\chi^{2}$ function for each distribution.
The fits yield a single-measurement precision of 0.8~\kms\ for Tillinghast + DS RVs, 0.7~\kms\ for MMT RVs and 
0.5~\kms\ for WIYN RVs.  See also Figure~\ref{fig:precV} for an analysis of the RV precision as a function of the stars' $V$ magnitudes.}
\label{fig:eoi}
\end{figure}

The majority of the CfA measurements come from the MMT and the Tillinghast Reflector (both using the Digital Speedometers).
We follow the same procedure as \cite{geller:08} to empirically determine the typical single-measurement precision 
for observations made at these two telescopes and at the WIYN Observatory with Hydra.
Specifically we fit a $\chi^2$ distribution of two degrees of freedom to the distribution of standard 
deviations of the first three RV measurements for each single-lined star in these samples, respectively.
The results for all stars observed $\geq3$ times at these telescopes are shown in Figures~\ref{fig:eoi}.
The dashed lines in each panel of Figure~\ref{fig:eoi} show the best fitting $\chi^2$ distribution functions, respectively,
which yield typical single-measurement precision values of 0.8~\kms\ for RVs from the Tillinghast Reflector + DS, 0.7~\kms\ for RVs from the MMT, 
and 0.5~\kms\ for RV from WIYN.
These single-measurement RV precision values for the CfA telescopes are in good agreement with those found by 
\citet{mathieu:86}, and also agree with the value found by 
\cite{hole:09} for observations of NGC 6819 taken at the MMT and the Tillinghast Reflector with the same instrumental setup.
Likewise, the single-measurement RV precision for observations at WIYN found here is similar to that found for the narrow-lined stars in
NGC 188, NGC 6819 and M35 observed with this same setup \citep{geller:08,geller:10,hole:09}.

\begin{figure}[!t]
%\epsscale{0.6}
\plotone{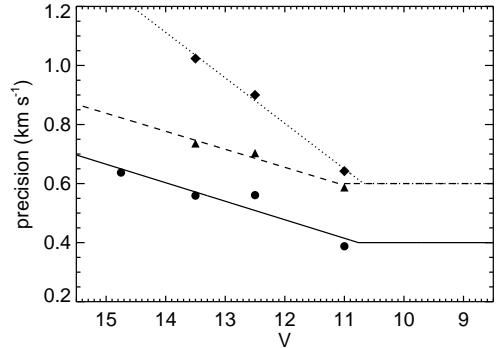}
\caption{\small Single-measurement RV precision as a function of $V$ magnitude for stars
observed at the WIYN telescope (circles and solid line), MMT (triangles and dashed line) and the 
Tillinghast Reflector + DS (diamonds and dotted line).  The points are placed at the centers of bins 
in $V$ magnitude (with edges at $V = $~15.5, 14, 13, 12, 10), and show results from our 
fits to $\chi^{2}$ distributions within these respective bins for each telescope.  The lines 
show linear fits to these data, with a floor of 0.4~\kms\ for WIYN RVs and 0.6~\kms\ for 
both the MMT and Tillinghast + DS RVs.}
\label{fig:precV}
\epsscale{1}
\end{figure}

For these three telescopes (WIYN, MMT and Tillinghast + DS), we also have sufficient data to characterize our RV precision as a function of 
$V$ magnitude (which, in general, correlates with the signal-to-noise of the measurement for a given integration time).  
As also found by \citet{geller:08} and \citet{mathieu:86} the single-measurement RV precision degrades towards fainter magnitudes.
In Figure~\ref{fig:precV} we show the single-measurement precision values resulting from $\chi^{2}$ distribution fits to observations from 
these three telescopes in bins of the stars' $V$ magnitudes.  
The lines show linear fits to these data.  To avoid unrealistic precision values, we impose a precision floor at 
the values in the bin containing the brightest stars, specifically at 0.4~\kms\ for WIYN RVs and 0.6~\kms\ for MMT and Tillinghast + DS RVs.  
We then use these fit lines to determine our single-measurement precision values for stars observed at these telescopes (given their respective
$V$ magnitudes) in later analyses.  

We also note that observations of the few rapidly rotating stars in our sample have poorer precision (see \citealt{geller:10}
for a detailed discussion of the effects of stellar rotation on the WIYN RV precision).  For these few stars, we do not attempt to derive
single-measurement RV precisions.  

We lack sufficient observations on the remaining telescopes to reliably utilize the $\chi^2$ distribution fitting technique.
For the Palomar Hale 5m RVs, the typical standard deviation of the measurements of a given star is 0.3 \kms\ \citep{mathieu:86};
we use this value as our single-measurement RV precision for stars observed at Palomar. 
For observations from the Wyeth 1.5m, there are only 12 stars with at least three RV measurements, and a third of these stars are in binaries (with orbital solutions).  
The mean of the standard deviations of the first three RVs for the remaining eight stars is 0.4~\kms, and we take this as an estimate of the 
single-measurement RV precision for observations from the Wyeth.
CORAVEL and TRES observations were primarily of binaries.  For CORAVEL RVs, we estimate the RV precision based on the $(O-C)$ residuals 
from the orbital fits in \citet{mathieu:90}, and find a typical precision of 0.5~\kms.

The correction of the TRES velocities for shifts in the instrumental
zero-point from month to month and year to year is now reliable at the
level of 0.02~\kms\ or better \citep[e.g. see][]{quinn:14}, much
better than the precision achieved for most of our TRES spectra of M67
stars due to SNR limitations set by photon noise.  Nevertheless, most
of our TRES observations of slowly-rotating single-lined stars in M67
do yield a velocity precison on the order of 0.1 (or perhaps 0.2)~\kms.
We assume a single-measurement RV precision of 0.1 \kms\ for TRES observations of
narrow-lined stars in the following analyses.

TRES observations of rapidly-rotating stars have proven to be much
more precise than the results for the same stars observed with the CfA
Digital Speedometers.  We attribute this to the higher SNR and better
wavelength coverage provided by TRES.  In particular, TRES has allowed
us to derive reliable RVs for two of the rapidly rotating BSS
in M67 that eluded success with our other facilities.

The long tails in the distributions extending beyond the best fitting curves shown in Figure~\ref{fig:eoi}
are populated by binary (and higher-order) systems. 
Moreover, because the standard deviation of the RVs from a binary should be significantly larger than the single-measurement RV precision,
we use the results of this analysis to identify binaries in our sample.
We discuss our criteria for identifying binaries in Section~\ref{sub:vvar}.

\section{The Combined CfA and WIYN Data Set}
\label{sec:data}

The Palomar velocities were adjusted to the native CfA velocity system
as described in \citet{mathieu:86}.  
Here we check for a possible zero-point offset between the WIYN and CfA data by comparing the mean RVs for non-RV variable stars
in both samples.  Specifically we select only those stars that have three or more observations in both samples,
with $P(\chi^2)>$0.01 and \eoi$<3$ (we discuss the derivation of these quantities in Section~\ref{sub:vvar}).
There are 95 such stars in our sample, and for each of these stars we calculate the mean RV from each sample.
The mean RVs for stars in the CfA sample that have observations at multiple telescopes are weighted by the respective
precision values discussed above.  We then calculate the star-by-star difference in the
mean RVs, defined as ${\rm RV_{CfA} - RV_{WIYN}}$, and find a mean difference of 0.008~\kms, with a standard error of the mean of 0.06~\kms.  
As this difference is well below the RV precision of any telescope used here (and less than the magnitude of the offset of the native CfA system 
from the IAU system, as discussed in Section~\ref{sec:obs}), 
we conclude that there is no significant zero-point offset at this level between the WIYN and CfA RVs.  

We therefore proceed in combining the two data sets without modification, and use this combined data
set for the following analyses.  In total there are \NRVC\ RVs of \NstarsC\ stars from the CfA and \NRVW\ RVs of \NstarsW\ stars from WIYN for a 
total of \NRV\ RVs of \Nstars\ stars in M67.

\begin{figure}[!t]
%\epsscale{0.5}
\plotone{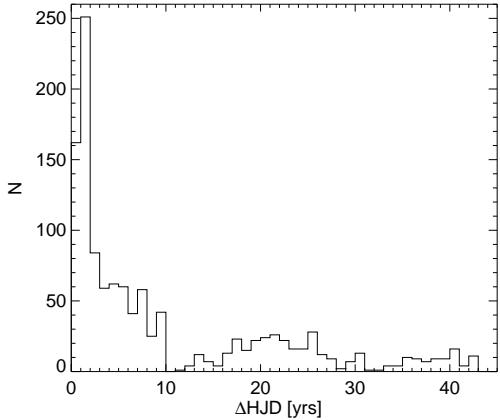}
\caption{\small Histogram of the maximum time span of RV measurements for stars in our M67 sample that 
have at least 2 observations.}
\label{fig:JDhist}
\end{figure}

In Figure~\ref{fig:JDhist} we plot a histogram of the maximum time span of RV measurements for the stars in our M67 sample (with at least 2 observations).  
For some stars, our observations span $>$40 years, and about one third of the stars in our sample have observations spanning at least 10 years.  
This long time baseline facilitates the detection of long-period binaries in our sample.  Again, we discuss our binary detection method in Section~\ref{sub:vvar}.

\subsection{Completeness of Spectroscopic Observations}
\label{sub:comp}

We define a ``primary stellar sample'' which extends from the brightest stars in the cluster to $V \leq 15.5$ and out to a 30 arcmin radius from the cluster
center.  In total, there are 903 stars in our primary sample.
We have at least one RV measurement for 902 of these stars, and $\geq$3 RV measurements for 899 of these stars.  
Moreover, there is only one star without any RV measurements;
2038 (S2312)\footnote{See Section~\ref{sec:results} for an explanation of our nomenclature for our IDs and cross-reference IDs.}   
is the second brightest star in our primary sample, is brighter than our WIYN sample, and 
is a \citet{sanders:77} proper-motion non-member.
The other three stars with $<3$ RV measurements are 1016 (S1306), 7033 (S1594), 7054 (F1295).
1016 is the brightest star in our sample and is a proper-motion non-member from three sources.
We have two observations of 1016 that are outside of the cluster RV distribution. 
We have observed 7033 and 7054 multiple times, and they both appear to be rapid rotators.
We have been unable to derive reliable RVs from the majority of these measurements.  
7033 has one proper-motion membership probability (from \citealt{sanders:77}) of 0\%.
7054 has no proper-motion measurements. Our one tentative RV measurement of 7054 is outside of the cluster RV distribution, but is uncertain due to the rotation.
In short, we have RVs for all likely cluster members within our complete sample.

\begin{figure*}[!t]
\epsscale{1.0}
\plotone{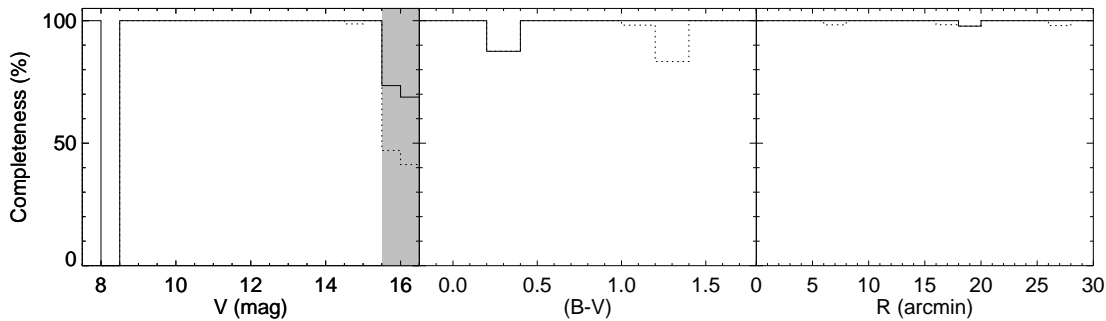}
\caption{\small Completeness histograms as a function of  
apparent $V$ magnitude (left), $(\bv)$ color (middle) and radius from the cluster center (right) for stars observed in M67.
In the $V$ completeness panel we show all stars observed in this survey.  The gray region shows
stars that are fainter than our primary sample selection of $V \leq 15.5$.
(Note that there are only two stars brighter than $V = 8.5$ in our sample, both of which are proper-motion non-members and 
neither of which have $\geq$3 observations, as discussed in the main text.)
In the $(\bv)$ and radius panels we show our completeness for stars in this primary sample only.
In all plots the percentage of stars observed $\geq$1 time is plotted with the solid line, 
while the percentage of stars observed $\geq$3 times is plotted with the 
dotted line.}
\label{fig:compltns}
\end{figure*}

Figure~\ref{fig:compltns} shows the percentage of stars with RV measurements in our stellar sample as functions of $V$ magnitude, $(\bv)$ color 
and distance from the cluster center. There is no significant trend in our completeness within the primary sample with magnitude, color or radius.  
We show our completeness for stars with $15.5 \leq V \leq 16.5$ in the left panel of Figure~\ref{fig:compltns} within the gray region. 
There are 418 stars within this magnitude range that are within 30 arcmin from the cluster center.  We have at least one RV measurement 
for 295 (71\%) of these stars, and $\geq$3 RV measurements for 182 (44\%) of these stars.  
As shown in Figure~\ref{fig:CfAsamp}, the CfA sample also contains observations of stars located at $>$30 arcmin from the cluster center.
We do not include these stars in our primary sample used in the subsequent analyses, but we include 
all observed stars in our RV summary table (Table~\ref{RVtab}, presented in Section~\ref{sec:results}).

\section{Results}
\label{sec:results}

In the following sections we analyze the RV measurements of each star in our stellar sample.  We first assess the RV variability 
of each star and then the membership probability. We use these quantities to classify stars with $\geq$3 observations
as cluster members or non-members, and RV variables (e.g., binaries) or non-variables (e.g., single stars).
We present the results of this analysis in Table~\ref{RVtab}, where we provide the WOCS ID ($ID_{\rm W}$)\footnote{We follow the same method as 
\citet{hole:09} to define WOCS ID based on the given star's $V$ magnitude and distance from the cluster center.}, a cross-reference
ID ($ID_{\rm X}$)\footnote{If available the cross-reference ID is taken from  \citet{sanders:77} and denoted by the prefix ``S''.
If there is no \citet{sanders:77} source match, we provide the \citet{montgomery:93} ID, if available, denoted by the 
prefix ``M''.  If both of those studies lack a source match, we provide the \citet{fan:96} IDs, if available, denoted by the prefix ``F''.
There are 13 sources in our table that do not have matches in these three references.
As with all sources, we provide their $RA$ and $Dec$ positions for matching to other catalogs.},
the J2000 right ascension ($RA$) and declination ($Dec$), the $V$ magnitude and $(B-V)$ color from 
\citet{montgomery:93}, where available\footnote{For the few stars in our sample that are not in the \citet{montgomery:93} survey
and for which we can find no other $BV$ photometry from the literature, we derive $V$ magnitudes from the 2MASS $JHK$ photometry 
using a similar relationship to \citet{girard:04}.  For these stars, we do not have $(\bv)$ colors.  
The stars with IDs that have a prefix of ``T'' are taken from \citet{mathieu:86} who use photometry from \citet{murray:68}.},
number of RV measurements in the WIYN ($N_{\rm W}$) and CfA ($N_{\rm C}$) samples (counting RVs for primary stars only), the modified Julian date (JD - 2400000) 
of the first observation ($JD_0$) and last observation ($JD_f$), the weighted mean RV ($\overline{\mathrm{RV}}$; for binaries with 
orbital solutions, we instead provide the center-of-mass, $\gamma$-RV) and
the weighted standard error of the mean RV ($RV_e$; for binaries with orbital solutions we instead provide the error on the $\gamma$-RV), 
the combined RV precision ($i$, defined as in \citealt{hole:09})
the RV membership probability (\PRV, defined in Section~\ref{sub:membership}), the proper-motion membership probabilities
from \citet[][$P_{\rm PMy}$]{yadav:08}, \citet[][$P_{\rm PMz}$]{zhao:93}, \citet[][$P_{\rm PMg}$]{girard:89} and \citet[][$P_{\rm PMs}$]{sanders:77}, where available, 
the \eoi\ and $P(\chi^2)$ values (defined in Section~\ref{sub:vvar}), the membership classification (see Section~\ref{sub:class}) and finally a
comment field.

We mark particularly notable stars within our stellar sample in this comment field of Table~\ref{RVtab}.
X-ray sources identified by \citet{belloni:98} with ROSAT are labeled with ``X'' followed by the source number given in their paper.  
Likewise X-ray sources identified by \citet{vandenberg:04} with Chandra are labeled with ``CX'' followed by the source number given in their paper.
Photometric variables 
\citep{gilliland:91,stassun:02,vandenberg:02,sandquist:03a,sandquist:03b,sandquist:03c,qian:06,bruntt:07,stello:07,pribulla:08,yakut:09}
are labeled with ``PV'', or ``PV?'' if the authors identify the photometric variability as uncertain 
(e.g., possible flare event detections, possible nearby source contamination, etc.).
We also label the W UMa's found in M67, and provide the GCVS names for photometric variables, where available (e.g., AH Cnc, ES Cnc, etc.).
Stars that are rotating significantly more rapidly than our instrumental resolution of $\sim$10 \kms\ are labeled as ``RR''.
Sources that are detected as triple systems are labeled with ``triple''.
Additionally, we label blue stragglers with ``BSS'', ``yellow giants'' with ``YG'', and the two sub-subgiants with ``SSG''. 
We briefly discuss a few of these notable stellar populations in Section~\ref{sub:snote}.

\subsection{Radial-Velocity Variability}
\label{sub:vvar}

We identify binaries in our M67 data using the \eoi\ statistic, which is the ratio of
the standard deviations ($e$) to the expected precision ($i$) of the RVs for a given star.
Members of binaries with orbital periods short enough to show significant RV variation in our data will have higher standard deviations
than expected for single stars.  Therefore high \eoi\ values indicate binaries or higher-order systems.
As a large number of stars in our sample have RV measurements
from multiple telescopes, and therefore multiple precision values, we use the formalism from \citet{bevington:92} to derive the 
$e$ and $i$ values for data with multiple precision values, as in \citet{hole:09}.

We derive an \eoi\ value for all single-lined stars with $\geq$3 RV measurements.  
Previous work by \citet{geller:08} shows that stars with \eoi$>$4 can be securely identified as binaries (or higher-order systems), 
and we initially followed this same cutoff for identifying binaries here.
In practice, however, we have derived RV orbital solutions for all proper-motion members in our primary sample 
with $3 \leq $~\eoi~$< 4$.
(There are only three proper-motion non-members in our primary sample with $3 \leq $~\eoi~$< 4$; each of these have mean RVs outside of the cluster distribution.)
Therefore here, we identify binaries in our data as having \eoi$\geq$3.

The uncertainties for RVs in double-lined (SB2) binaries 
are not well established, and therefore we do not derive \eoi\ values for these stars.  Instead we identify SB2s as binaries directly by inspection of
the spectra and the morphology of the peaks of the cross-correlation functions.
(We derive the uncertainties on the mean RVs, $RV_e$, for SB2s without orbital solutions using the measurement precisions that we derive above for single-lined stars; 
we suspect that the RV uncertainties for such SB2s quoted here are in fact lower limits.)

In addition to the \eoi\ statistic, we also provide the $P(\chi^2)$ value.  This statistic tests the hypothesis that
a given star's distribution of RVs is consistent with a constant value at the mean RV.  Members of binaries with short 
enough orbital periods to show significant RV variations in our data will be inconsistent with a constant RV and therefore will 
have small $P(\chi^2)$ values.
To derive the $P(\chi^2)$ value, we first calculate the standard $\chi^2$ statistic,
using the weighted mean RV and the respective
precision of each RV measurement (discussed in Section~\ref{sub:prec}).  For each observed 
star, we use the number of RV measurements minus one for the degrees of freedom. Then $P(\chi^2)$ is the 
corresponding probability for obtaining a value of $\chi^2$ greater than or equal to the observed value
with the given degrees of freedom.   
Again, we only derive $P(\chi^2)$ values for single-lined stars with $\geq$3 measurements.  

In the following analysis we will identify binaries by having \eoi$\geq3$, or having a binary orbital solution.
(The additional $P(\chi^2)$ statistic is provided for interested readers who prefer to use these values to perform their own selection of binaries.)
All stars with \eoi~$<3$ for which we have not derived binary orbital solutions, are labeled as ``single''.
However,  some of these stars are undoubtedly in long-period binaries, currently beyond our detection limit.

Finally, we also identify five likely triple stars in our sample, and label them as such in Table~\ref{RVtab}.  
In most cases the systems are double lined, where the RVs for one star vary on a much shorter time scale 
than those of the other star.  For a few, we see hints of tertiary velocities at low signal-to-noise, but we have yet to 
analyze these spectra in detail to derive all three velocities simultaneously. 
Additional triple stars may be detectable in our sample, for instance within the RV residuals from binary orbital solutions, but we save 
this analysis for a future paper.

\subsection{Membership}
\label{sub:membership}

\begin{figure*}[!t]
\epsscale{1.0}
\plottwo{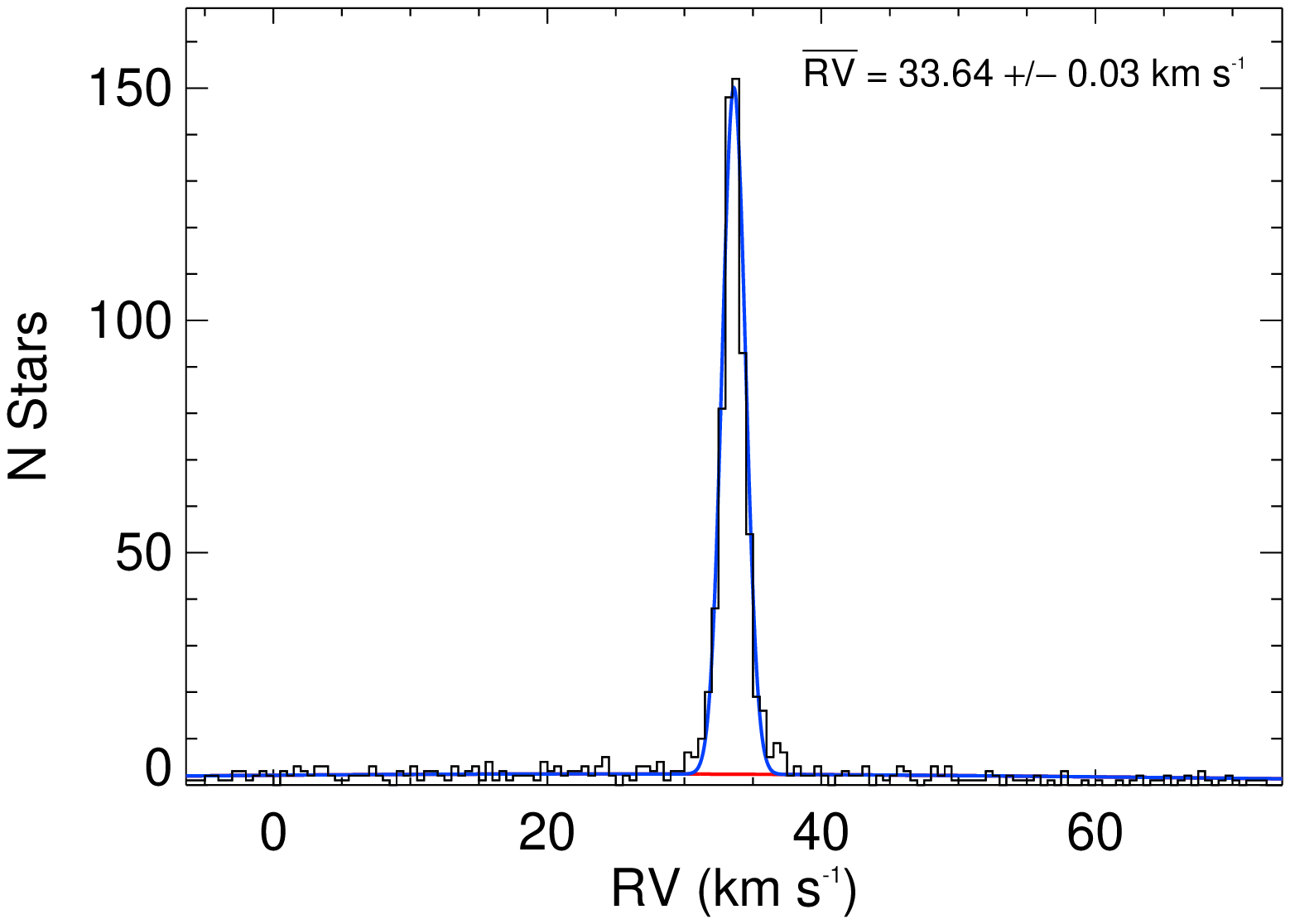}{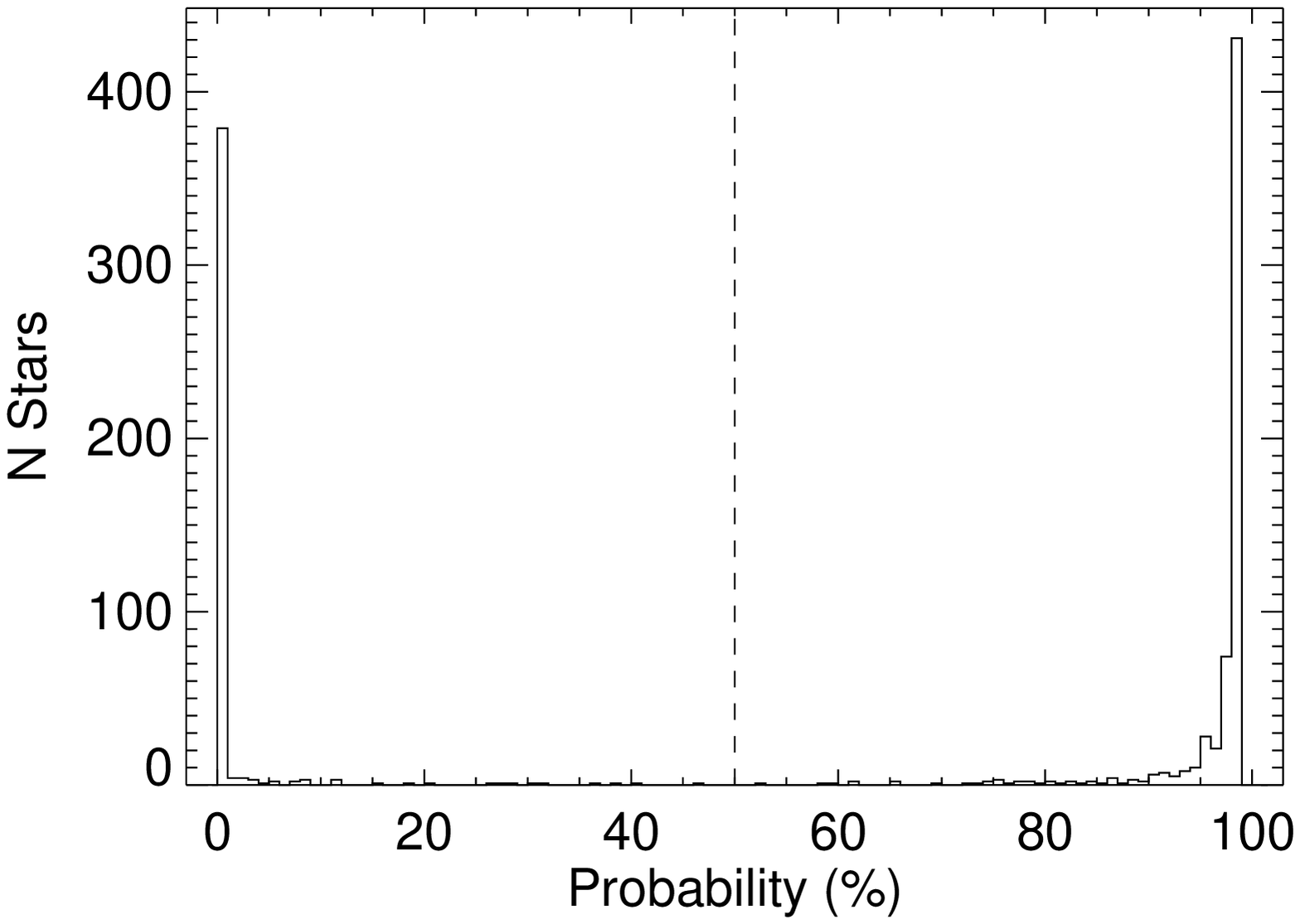}
\caption{\small  RV distribution (left) and the distribution of RV memberships (right) for stars in our M67 stellar sample.  
To derive the RV distribution we use the weighted mean RVs for stars with $\geq$3 observations with
$\eoi < 3$, and $\gamma$-RVs for binaries with orbital solutions.  (Here we exclude
detected binaries without orbital solutions, because their $\gamma$-RVs are unknown.)
We use bins of 0.5 \kms.
The red line shows the field component to our simultaneous two-Gaussian fit to the observations, and the 
blue line shows the combined fit of the cluster and field distributions, 
which we use to derive the membership probabilities for observed stars in M67.
On the right, we show a histogram of RV memberships (rounded to the nearest percent) of the same stars used in constructing
the RV distribution.
RV members are defined as having \PRV$ > 50$\%.}
\label{fig:mem}
\end{figure*}

\begin{table}[!t]
\centering
%\begin{minipage}{0.45\textwidth} %silly fix, but at least this works
\caption{\label{t:fit}Gaussian Fit Parameters For Cluster and Field Radial-Velocity Distributions \vspace{0.2em}}
%\end{minipage}
\begin{tabular}{c r@{\hspace{0.5em}}c@{\hspace{0.5em}}l r@{\hspace{0.5em}}c@{\hspace{0.5em}}l}
\hline
\hline
\relax\\[-1.7ex]
Parameter & \multicolumn{3}{c}{Cluster} & \multicolumn{3}{c}{Field} \\
\relax\\[-1.7ex]
\hline
\relax\\[-1.7ex]
Ampl. (Number)                  &  147.8  & $\pm$ & 0.7      & 2.38  & $\pm$ & 0.10 \\
$\overline{\mathrm{RV}}$ (\kms) &  33.615 & $\pm$ & 0.005    & 23.4  & $\pm$ & 2.2 \\
$\sigma$ (\kms)                 &  0.854  & $\pm$ & 0.005    & 46.7  & $\pm$ & 2.2 \\
\relax\\[-1.7ex]
\hline
\relax\\[-1.7ex]
\end{tabular}
\end{table}

In Figure~\ref{fig:mem} we show the distribution of weighted mean RVs for all stars 
in our sample with $\geq$3 RV measurements and \eoi~$< 3$,
and $\gamma$-RVs for binaries with orbital solutions.
For stars with RVs from multiple telescopes, we use the respective RV precision values to 
calculate the weighted means, and use those results here.
The cluster RV distribution is clearly distinguished from that of the field as the 
narrow distribution peaked at a mean RV of $+33.6$~\kms.  

In order to derive membership probabilities, we first fit simultaneous one-dimensional 
Gaussian functions to the cluster and field RV distributions, $F_{c}(v)$ and $F_{f}(v)$, respectively.
(We exclude from this fit binaries without orbital solutions, as their $\gamma$-RVs are unknown.)
The resulting combined fit to the cluster and field distributions are shown in the blue line in the left panel of Figure~\ref{fig:mem}, 
and the fit parameters are given in Table~\ref{t:fit} (fit to a histogram with a bin size of 0.5 \kms).
We then use the following equation,
\begin{equation} \label{memeq}
P_{\rm RV}(v) = \frac{F_{c}(v)}{F_{f}(v)+F_{c}(v)}  
\end{equation}
\citep{vasilevskis:58} to calculate the RV membership probability \PRV($v$) for a given star in our sample.  

We only compute membership probabilities for stars with $\geq3$ observations.
For non-RV-variable stars, we use the weighted mean RV in the calculations.
For binaries with orbital solutions, we use the $\gamma$-RV.  
For RV-variable stars without orbital solutions, we cannot calculate a reliable RV membership probability, as the $\gamma$-RV is unknown.  
For these stars we instead provide a preliminary membership classification, described in Section~\ref{sub:class}.

We plot the resulting distribution of RV membership probabilities in the right panel of Figure~\ref{fig:mem}.
Cluster and field stars are cleanly separated, and
we choose a cutoff of \PRV$\geq$50\% to define our M67 cluster member sample.  
Using our sample of single cluster members with \eoi~$< 3$ and binary 
cluster members with orbital solutions, we find a mean cluster velocity of $+33.64$~\kms\ (with an internal precision of $\pm 0.03$~\kms), 
in good agreement with previous RV surveys \citep[e.g., see][and references therein]{yadav:08}.

As stated in Sections~\ref{sec:obs}~and~\ref{sec:data}, the RVs in this paper are all on the native CfA system,
which is found to be shifted by $-0.14$~\kms\ compared to the IAU system.  Therefore, 
the mean cluster velocity quoted here may also have this offset compared to the IAU system 
(i.e., $+0.14$~\kms\ may need to be added to our velocities to get onto the IAU system).

The ratio of the areas under the Gaussian fits to the cluster and field distributions provides an 
estimate of the field star contamination.  
At a membership probability of 50\%, we expect a 3.5\% contamination from field stars in our RV cluster member sample (i.e., $\sim20$ stars) .

\subsubsection{Comparison of Radial-Velocity and Proper-Motion Membership Probabilities}
\label{ssub:cpm}

Here we compare with four published proper-motion membership catalogs for M67: 
\citet{sanders:77}, \citet{girard:89}, \citet{zhao:93} and \citet{yadav:08}.  For this comparison we use only non-RV variable stars and binaries 
with orbital solutions in our primary sample (thereby excluding binaries whose $\gamma$-RVs are unknown) so as to ensure secure 
RV membership probabilities.  There are 836 stars in our primary sample that meet these criteria.  

We find excellent agreement with all four proper-motion sources.
There are 456 stars in this sample with \citet{sanders:77} proper-motion membership probabilities $\geq$50, of which 
430 (94\%) also have RV membership probabilities $\geq$50\%.  
410 stars in this sample have \citet{girard:89} proper-motion
membership probabilities $\geq$50\%, of which 396 (97\%) also have RV membership probabilities $\geq$50\%.  
164 stars in this sample have \citet{zhao:93} proper-motion membership probabilities $\geq$50\%, of which
149 (91\%) also have RV membership probabilities $\geq$50\%.  
Finally, 330 stars in this sample have \citet{yadav:08} proper-motion membership probabilities $\geq$50\%, of which
200 (91\%) also have RV membership probabilities $\geq$50\%.  
If we examine stars in this sample that have
proper-motion membership probabilities of $\geq$50\% from all four references, we find that 95/100 (95\%) also 
have RV memberships $\geq$50\%.

For stars with both RV and proper-motion membership probabilities, we combine the results to further refine our 
cluster member sample.  We first take all stars that have RV membership probabilities $\geq$50\% 
as candidate cluster members. We then examine the available proper-motion membership probabilities for each star to remove 
non-members from this sample.  For each proper-motion reference we assign the same cutoff of $\geq$50\% membership 
probability to denote cluster members.  If the available proper-motion membership values for a given star all
indicate that the star is a non-member, we remove it from our cluster member sample.
For stars with proper-motion membership probabilities from multiple references
that disagree on membership status, we allow each proper-motion reference one ``vote'' on membership (either member or non-member) 
and take the majority vote to determine the proper-motion membership status.
If this procedure results in a tie, we instead use the result from the highest precision proper-motion membership 
survey for the given star to determine membership (in the order of \citealt{girard:89}, \citealt{yadav:08}, \citealt{zhao:93}, 
then \citealt{sanders:77}, with \citealt{girard:89} at the highest precision).  

In practice, only 24 stars in our primary sample have proper-motion membership probabilities that result in a ``tie'', all but two of which 
have only two proper-motion membership values.  Also for reference, there are 35 stars in our primary sample with proper-motion memberships 
from all four studies where one disagrees with the other three.  
In 23 of these cases, \citet{zhao:93} is the outlier.
In 8 of these cases, \citet{yadav:08} is the outlier.    
In 4 of these cases, \citet{girard:89} is the outlier.

Of the 515 stars in our primary sample that have secure RV memberships $\geq$50\% (i.e., SM and BM stars, see Section~\ref{sub:class}) and 
proper-motion measurements, 455 pass this proper-motion membership test, and are therefore deemed bona fide cluster members.  
This $\sim$12\% contamination of our RV member sample is about 3.5 times larger than we estimate 
above based solely on the RV distributions of the cluster and field stars.  We note that 75\% of these RV members that we determine
to be proper-motion non-members from the algorithm above reside outside of 15 arcmin from the cluster center.
Also about 23\% of these RV members that are proper-motion non-members have at least one proper-motion membership value 
(from one of the references above) greater than 50\%.  
However, to minimize possible field star contamination, we remove all of these proper-motion non-members from our cluster member sample.

\subsection{Membership Classification of Radial-Velocity Variable Stars}
\label{sub:class}

\begin{table*}[!t]
\centering
%\begin{minipage}{0.75\textwidth} %silly fix, but at least this works
\caption{\label{t:class}Description of Stars or Star Systems Within Each Membership Class \vspace{0.5em}}
%\end{minipage}
\begin{tabular}{cp{0.21\linewidth}p{0.59\linewidth}c}
\hline
\hline
\relax\\[-1.7ex]
Class & Description & Criteria & Number \\
\relax\\[-1.7ex]
\hline
\relax\\[-1.7ex]
SM & Single Member &     $\geq$3 RVs, \eoi~$< 3$, \PRV$(\overline{\rm RV}) \geq 50$\% AND \PPM$ = $~M  & \SM \\
SN & Single Non-member & $\geq$3 RVs , \eoi~$< 3$, \PRV$(\overline{\rm RV}) < 50$\% OR \PPM$ = $~NM & \SN \\
BM & Binary Member &     binary orbit, \PRV$(\gamma$-RV$) \geq 50$\% AND \PPM$ = $~M & \BM \\
BN & Binary Non-member &  binary orbit,  \PRV$(\gamma$-RV$) < 50$\%  OR \PPM$ = $~NM & \BN \\
BLM & Binary Likely Member & $\geq$3 RVs , \eoi~$\geq 3$, \PRV$(\overline{\rm RV}) \geq 50$\% AND \PPM$ = $~M & \BLM \\
BU &  Binary with Unknown RV Membership & $\geq$3 RVs, \eoi~$\geq 3$, \PRV$(\overline{\rm RV}) < 50$\% AND the range in RV measurements includes the cluster mean RV AND \PPM$ = $~M  & \BU \\
BLN & Binary Likely Non-Member & $\geq$3 RVs, \eoi~$\geq 3$, \PRV$(\overline{\rm RV}) < 50$\% AND all RV measurements are either at higher or lower RV than the cluster distribution, OR \PPM$ = $~NM  & \BLN \\
U & Unknown RV Membership & $<3$ RVs & \U \\
\relax\\[-1.7ex]
\hline
\relax\\[-1.7ex]
\end{tabular}
\end{table*}

For stars that show no significant RV variability and binaries with orbital solutions, we can calculate precise RV membership 
probabilities that allow us to separate cluster members from field stars, as described above.  However, 
for binary stars that do not have RV orbital solutions, we cannot calculate reliable RV membership probabilities 
because we do not know their $\gamma$-RVs.  Thus we follow a similar method to \citet{geller:08,geller:10} and \citet{hole:09} in order 
to provide a qualitative classification of the membership and variability for each observed star.  

Our classification scheme is defined in Table~\ref{t:class}, where we identify our eight qualitative membership classes, 
the selection criteria for each class, and also the number of stars that reside in each class. 
The selection criteria depend on the number of RVs, the \eoi\ value, the RV membership probability (\PRV, 
determined either using the mean RV, $\overline{\rm RV}$, or the $\gamma$-RV, for 
binaries with orbital solutions), and the proper-motion membership ``vote'' described in Section~\ref{ssub:cpm}.
For the proper-motion membership vote, we use ``\PPM$ = $M'' to indicate proper-motion members, and ``\PPM$ = $NM''
to indicate proper-motion non-members.  If a star does not have a proper-motion measurement, we use only the RV criteria to 
classify the star.

In short the single members/non-members (SM/SN) or binary members/non-members (BM/BN) are stars with secure 
membership status; these are the only stars for which we can provide reliable RV membership probabilities.
For RV variable stars with $\geq$3 RV measurements that do not have a binary orbital solution,
we divide our classification into three groups, including binary likely members (BLM), 
binaries with unknown RV membership (BU) and binary likely non-members (BLN).  We anticipate that 
eventually orbital solutions derived for BLM binaries will place them in the BM category, while 
those for the BLN binaries will place them in the BN category (since many of these sources are proper-motion non-members, and 
for those that are not, it is unlikely that an orbital solutions will place their $\gamma$-RVs within the cluster distribution).
Binaries with unknown RV membership (BU's) are proper-motion members.  Therefore here we assume that these
are indeed cluster members (unlike in other WOCS papers, where proper-motion memberships were unavailable). 
Stars with $<$3 RV measurements have unknown RV membership (U), as these stars do not meet our 
minimum criterion for deriving RV memberships or \eoi\ values.
In the following analyses we restrict our cluster member sample to only include the \NMEM\ stars
classified as either SM, BM, BLM or BU.

Finally, as mentioned above, stars that show broadened spectral features (e.g., due to rapid rotation) do not have secure 
single-measurement RV precision values, and therefore in most cases we cannot confidently classify such stars as binaries or singles.  
We provide our best assessment of the binarity of these sources in Table~\ref{RVtab} and indicate our uncertainty in their class with 
parentheses, e.g. (BL)M, (S)N, etc.

\section{Discussion}
\label{sec:disc}

In the following section, we use our confirmed cluster members to investigate the CMD (Section~\ref{sub:CMD}), identify and discuss a few 
notable stellar populations (Section~\ref{sub:snote}), analyze the spatial distribution of the single, binary, giant and BSS cluster populations
(Section~\ref{sub:spatdist}) and derive the velocity dispersion of the solar-type stars in our cluster member sample (Section~\ref{sub:vdisp}).

\subsection{Color-Magnitude Diagram}
\label{sub:CMD}

\begin{figure*}[!t]
\epsscale{1.0}
\plottwo{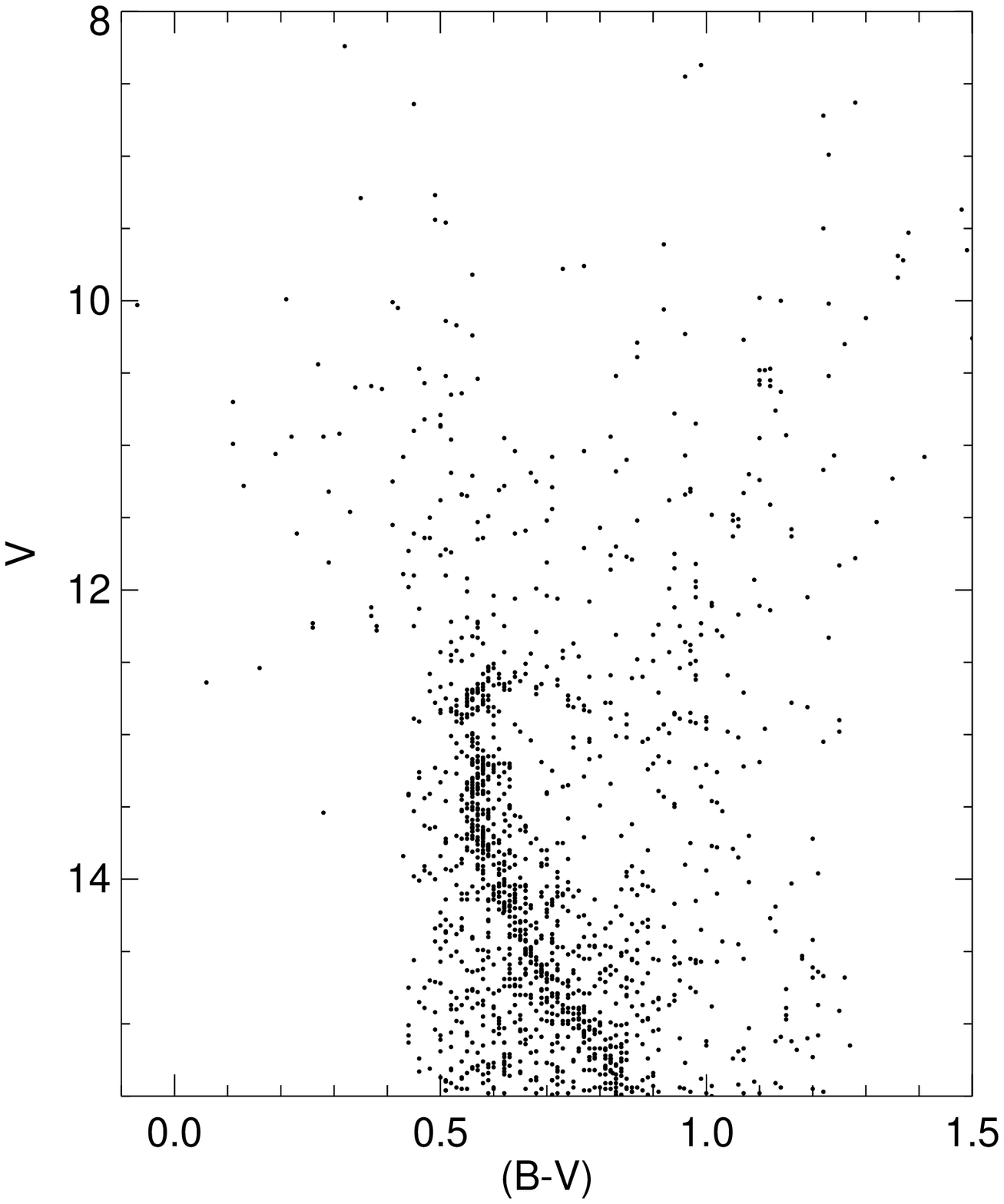}{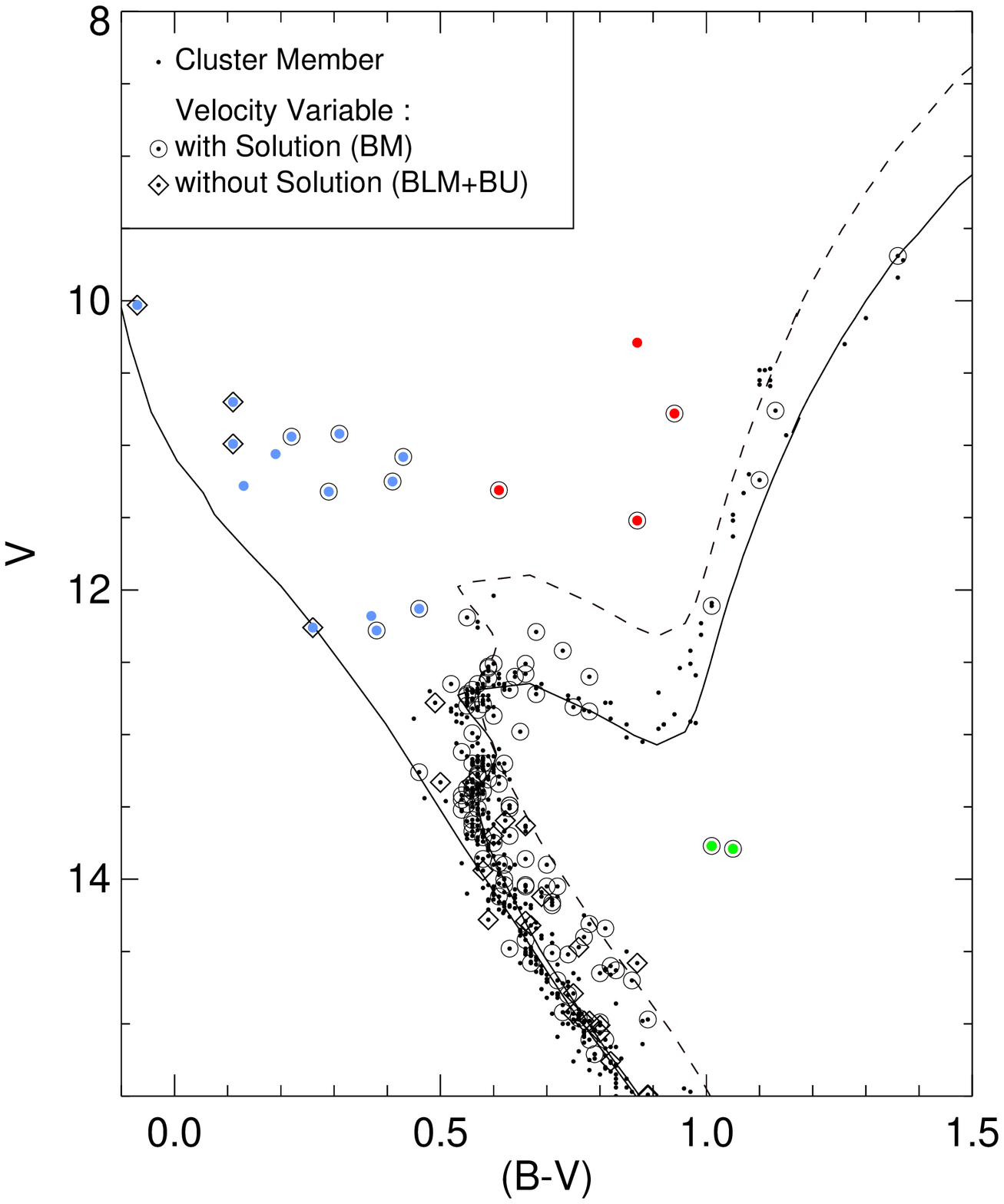}
\caption{\small  Color-magnitude diagrams for all stars in our M67 stellar sample 
with $V<15.5$ (and available $(\bv)$ colors; left) and only cluster members within the same magnitude range (right).  
We take here as cluster members all stars that reside in the SM, BM, BLM, and BU classes, as well 
as stars in the U class and rapid rotators that are proper-motion members (see Section~\ref{sub:class}).
Binary members with orbital solutions (BMs) are 
circled, and velocity variables without orbital solutions that are likely cluster members (BLMs and BUs) are shown in 
diamonds. For comparison, in the right panel we also plot a zero-age main-sequence 
isochrone and a 4 Gyr isochrone \citep[solid lines][]{marigo:08} using 
$(m-M)_V= 9.6$, $E(B-V)=0.01$ and solar metallicity.
We also plot the equal-mass binary locus (dashed line) obtained by shifting 
the 4 Gyr isochrone by -0.75 mag.}
\label{fig:cmd}
\end{figure*}

In Figure~\ref{fig:cmd} we plot the CMD of all stars in our stellar sample with $V<15.5$ (and available $(\bv)$ colors; left) 
and only the confirmed cluster members within the same magnitude range (right).  
Without removing non-members, the main-sequence of the cluster is visible, but the BSS and giant populations
cannot be distinguished from the field.  Applying both our proper-motion and RV membership criteria reveals a rich cluster 
containing well populated main-sequence, subgiant and giant branches as well as a large population of BSS (blue points), four yellow giants (red points),
and two sub-subgiants (green points).  Specifically, we plot here stars that we classify as SM, BM, BLM and BU, and we also include here stars in 
the U category as well as rapid rotators that are proper-motion members.

For comparison, we also plot a 4 Gyr isochrone and a zero-age main-sequence isochrone (solid lines), as well as an 
equal-mass binary line (dashed line).  
The isochrones are from \citet{marigo:08}, and use solar metallicity, $(m-M)_V= 9.6$ and $E(B-V)=0.01$,
consistent with recent results derived in the literature (see Section~\ref{sec:intro}).
We include the isochrones simply to help guide the eye; they are not meant as a fit to the observed data.

Binaries with orbital solutions are circled, and binaries without orbital 
solutions are marked with diamonds.  
As is seen clearly here, and also noted by \citet{latham:96} and \citet{latham:07}, the ``exotic'' stars (i.e., the BSS, yellow
giants, and sub-subgiants) have a remarkably high binary frequency.  
In total we identify 14 BSS, four yellow giants (one of which is outside of a 30 arcmin radius from the cluster center) and two sub-subgiants in our stellar sample. 
At least 16/20 (80\%~$\pm$~20\%) of these exotic stars are RV variables (and others may also be binaries, e.g.\ with long-period orbits that are currently 
outside of our detection limit).  In comparison, 122/538 (22.7\%~$\pm$~2.1\%) of the ``normal'' stars, located in more typical positions in the CMD, show RV variability 
indicative of binarity.  Thus the exotic stars have a significantly higher frequency of binaries than the normal stars.  This result is 
similar to that found in the old (7 Gyr) open cluster NGC 188, where \citet{mathieu:09,mathieu:15} find that 80\% of the NGC 188 BSS have 
binary companions (roughly three times the binary frequency of the main-sequence stars in NGC 188).  

We also note the very tight red giant sequence, first discussed in detail by \citet{janes:84}.  Such a tight red giant
sequence is expected for a single coeval population, but, interestingly, the giants of NGC 188 
\citep[e.g.][]{geller:08} and the intermediate-age (2.5 Gyr) open cluster NGC 6819 \citep[e.g.][]{hole:09} show a much larger scatter than the giants in M67.  
The origin of the scatter in these other open clusters is unknown.

\subsection{Stars of Note}
\label{sub:snote}

Within our cluster member sample, there are a number of intriguing stellar populations, from those that lie far from the predicted locus of single 
stars from stellar evolution theory (including the well known BSS, Section~\ref{ssub:BSS}, and anomalous giants, Section~\ref{ssub:AG}), to those 
that have physical characteristics very similar to those of our Sun (Section~\ref{ssub:stwins}), and even three exoplanet host stars (Section~\ref{ssub:planet}). 
In the following we briefly identify and discuss these stars of note.

\subsubsection{Blue Stragglers}
\label{ssub:BSS}

M67 is home to one of the most well-studied BSS populations
 \citep[see, e.g.,][]{eggen:81,peterson:84,mathieu:86,manteiga:89,manteiga:91,gilliland:92,leonard:92,milone:92,milone:94,leonard:96,landsman:97,landsman:98,deng:99,shetrone:00,hurley:01,vandenberg:01,sandquist:03c,chen:04,hurley:05,sandquist:05,zhang:05,andronov:06,tian:06,bruntt:07,latham:07,liu:08,pribulla:08,lu:10}.
In total, we identify 25 candidate BSS that are discussed in the literature: 
1006 (S1066),
1007 (S1284),
1010 (S977),
1017 (S1466), %NM
1020 (S751), %blue hook
1025 (S1195),
1026 (S1434),
2007 (S984), %turnoff
2008 (S1072), %YG
2009 (S1082),
2011 (S968),
2013 (S1267),
2015 (S792), %turnoff
2068 (S277), %outside of 30 arcmin, and also near turnoff
3005 (S1263),  
3009 (S1273), %turnoff
3010 (S975),
3013 (S752),
4003 (S1036), %blue hook
4006 (S1280),
5005 (S997),
5071 (S145), %NM by RV
6038 (S2226),
8006 (S2204), %blue hook
9005 (S1005). %blue hook

We find 1017 and 5071 to be cluster non-members based on kinematic membership information,
and therefore we remove these from the list of BSS. 1017 is a non-member by both proper motions and RVs.
5071 has a proper-motion membership probability of 94\% from \citet{sanders:77}, but the star appears to be single with a mean RV or 
$39.79 \pm 0.12$~\kms, which results in a 0\% RV membership probability.  

By examination of the CMD shown in Figure~\ref{fig:cmd}, we remove an additional nine stars from the BSS sample.
1020, 4003, 8006 and 9005  lie close to the blue hook of the cluster (given the \citealt{montgomery:93} BV photometry), 
and therefore we conservatively exclude these stars from our BSS sample.  
2007, 2015, 2068 and 3009 are above the turnoff, but reside in a region expected to be populated by binaries containing normal main-sequence turnoff stars.
We only detect a binary companion to 2068.  However, the remaining three stars may have long-period companions, currently beyond our detection limit, with 
high enough masses to explain their location on the CMD.  We therefore do not include these stars in our BSS sample.
2008 is the reddest BSS candidate in this sample, residing $\sim$1.5 magnitudes directly brighter than the main-sequence turnoff. 
The BSS status of 2008 is somewhat ambiguous, as it may currently be evolving toward the giant branch. 
We choose to exclude 2008 from our BSS sample and will 
include it in our ``anomalous giant'' sample discussed below. (\citealt{mathieu:86b} also exclude 2008 from their BSS 
sample based on similar arguments.)

Thus in total we find M67 to have 14 BSS in our primary stellar sample.
Remarkably, 11 of these 14 BSS show significant RV variations indicative of binary companions.
Seven of these BSS have secure orbital solutions, and an additional two rapidly rotating BSS (1026, 4006) have preliminary orbital solutions.
The detected binary frequency amongst the M67 BSS is 79\%~$\pm$~24\%.  This high binary frequency for the M67 BSS is similar to that
found in the old (7 Gyr) open cluster NGC 188 \citep{mathieu:09}, and in the field \citep{carney:01}.
We will discuss these BSS in detail, including their binary properties, in a subsequent paper.

\subsubsection{Anomalous Giant Stars}
\label{ssub:AG}

There are six stars with $(B-V)$ colors consistent with the subgiant or giant branches but have magnitudes and/or chemical abundances that set the apart from 
the more typical giants along the isochrone shown in Figure~\ref{fig:cmd}.  We discuss these anomalous giant stars below.

\textbf{Sub-subgiants: }15028 (S1113) and 13008 (S1063) are fainter than the subgiant branch but redder than the main-sequence.  
These sub-subgiants are both members of binary systems and are also X-ray sources.  
Their origins are unknown, and we refer the reader to \citet{mathieu:03} for a very detailed discussion 
on the available observations for these stars.

\textbf{Lithium-rich Subgiant: }\citet{canto:06} find the subgiant 6008 (S1242) to have an anomalously high Lithium abundance as compared to normal main-sequence 
turnoff stars in the cluster, although it falls within the normal subgiant branch on the CMD.  
We confirm the membership of 6008, and we also confirm that 6008 is in a binary.

\textbf{Yellow Giants: }1015 (S1237), 1112, 2002 (S1040) and 2008 (S1072) are all brighter than the giant branch and have often been referred 
to as ``yellow giants''.  1112 is 55.6 arcmin from the cluster center, 
and is therefore not included in our primary stellar sample.  This star shows no evidence for a binary companion.
The other three yellow giants are all members of binaries.
2002 was studied in detail by \citet{landsman:97,landsman:98} who find the secondary to be a low-mass He white dwarf, suggesting that 
the system went through an episode of mass transfer while the donor was on the giant branch.  Therefore 2002 may have been a BSS
in the recent past, and is now in the process of evolving towards the giant branch.  
2008 may also have previously been a BSS that is now evolving towards the giant branch \citep{mathieu:86b}.  A similar evolutionary scenario
may explain the anomalous CMD location of 1015.

\subsubsection{Solar Twins}
\label{ssub:stwins}

M67 is of a similar age and chemical composition to the Sun, and is therefore an ideal target for investigating solar analogs 
\citep{pasquini:08,reiners:09,giampapa:06,castro:11,onehag:11}.  
Stars 7003 (S1041), 10018 (S1462), 11018 (S1095), 12012 (S996), 13021 (S945), 14014 (S779), 16011 (S770), 16023 (S2211), 17026 (S1335), and 18013 (S785)
were found to have effective temperatures consistent with the Sun, and are the closest analogs to the Sun in M67.  We confirm that these 10 stars 
are cluster members, and all but one appear to be single (with 10018 at \eoi$=$4.19).  
16023 was found to be a photometric variable by \citet{stassun:02}, although variability
was only detected above the $3\sigma$ level in the $B$ band (and not in $V$ or $I$).  We find 16023 to have \eoi$=$0.93.

\subsubsection{Exoplanet Host Stars}
\label{ssub:planet}

\citet{brucalassi:14} identify three stars hosting roughly Jupiter-mass exoplanets in M67: 1045 (S364), 13014 (S802) and 16011 (S770).  Our measurements are not sensitive 
enough to detect such low-mass companions, but we confirm that none appear to have stellar-mass companions and that all three are cluster members.
We also note that 1045 is an X-ray source (X19 from \citealt{belloni:98}).

\subsection{Radial Distribution of Cluster Members}
\label{sub:spatdist}

\begin{figure}[!t]
%\epsscale{0.6}
\plotone{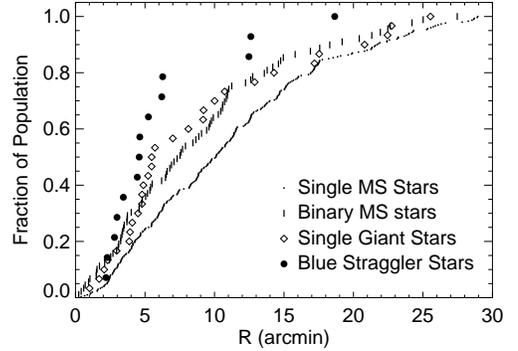}
\epsscale{1.0}
\caption{\small Cumulative projected radial distribution of cluster members in M67. We divide our primary 
sample into single and binary main-sequence (MS) stars, single giants and 
blue stragglers.  The giants are identified as having $(B-V) > 0.9$, and we exclude the anomalous giants discussed
in Section~\ref{sub:snote}.  The blue straggler sample is identified as discussed in Section~\ref{sub:snote} and 
shown in Figure~\ref{fig:cmd}.  Both the binaries and blue stragglers are significantly centrally concentrated with 
respect to the single main-sequence stars. }
\label{fig:spatdist}
\end{figure}

The spatial distribution of stars in M67 has been studied in detail \citep[e.g.][]{mathieu:86b,zhao:96,sarajedini:99,bonatto:03,davenport:10}.
M67 is mass segregated, as is expected for a 4 Gyr cluster with 
a half-mass relaxation time of 100 Myr \citep{mathieu:86b}.
In light of our new RV memberships and identification of binaries, 
we briefly re-investigate the spatial distribution of cluster member populations using our primary sample.
In Figure~\ref{fig:spatdist} we show the projected radial distribution for the single main-sequence stars,
binary main-sequence stars, single giants and BSS.  Singles are stars classified as SM, and binaries are
classified as either BM, BLM or BU.

A Kolmogorov-Smirnov (K-S) test shows that the main-sequence binaries are centrally concentrated with 
respect to the main-sequence single stars at the 98.7\% confidence level. Comparing all single stars to all binaries shows 
that the binaries are more centrally concentrated at the 99.8\% confidence level. 
Because the total mass of a given binary is generally more massive than that of a single star in our primary sample, 
this results confirms that M67 is mass segregated.

\citet{mathieu:86b} find the BSS to be centrally concentrated with respect to the single stars near the cluster turnoff.
Our BSS sample is slightly different than that of \citet{mathieu:86b}. 
Nonetheless, the result is the same.
A K-S test comparing the BSS to the single main-sequence stars show that the BSS are centrally concentrated at the 99.7\% confidence level.
This result suggests that the BSS are more massive than the main-sequence stars in the cluster, as was noted by \citet{mathieu:86b}.

In many globular clusters and also the old open cluster NGC 188, BSS show a bimodal projected radial distribution
\citep[e.g.][]{ferraro:97,ferraro:12,mapelli:06,geller:08}.  We do not observe evidence for a bimodal radial distribution 
in our M67 BSS sample.  However, our sample only extends to 6 or 7 core radii.
In NGC 188, which is of a similar 
dynamical age to M67, the halo BSS population begins at roughly 6 core radii and extends to about 13 core radii (at least).
Thus if M67 and NGC 188 have similar BSS radial distributions, we would only expect to see the inner population of M67 in 
our current sample, and would require a survey extending to roughly twice the current radial extent to search for a bimodal structure.

Interestingly, the giants appear to follow a similar spatial distribution as the binaries, despite having very similar masses to the upper main-sequence stars
included in our primary sample.
However, given the relatively small sample size of the single giants, a K-S test comparing the spatial 
distributions of the single giants and single main-sequence stars returns a distinction at only the 95.9\% confidence level.
For comparison, the NGC 188 giants do not follow the distribution of the binaries, and instead follow closely to the single cluster 
members \citep[see Figure 9 in][]{geller:08}, both when considering the entire spatial extent of the RV survey and when 
limiting to a similar number or core radii as our M67 sample.
This difference is intriguing, but the small sample sizes of the giants in both clusters make such comparisons uncertain.

\subsection{Cluster Radial-Velocity Dispersion and Virial Mass}
\label{sub:vdisp}

We follow the method of \citet{geller:10} to calculate the RV dispersion of the M67 single main-sequence, subgiant and giant members in our primary sample.
(We exclude the binaries and BSS because these have significantly different spatial distributions, likely due to their higher masses.)
This method assumes that the observed RV dispersion is composed of two components, one which we will call the ``combined RV dispersion'' and 
one for the observational error, and also assumes that the distributions of RVs and errors are Gaussian.
The observed dispersion is directly measured, and by subtracting off the component from observational
error, we recover the combined RV dispersion. The combined RV dispersion itself contains contributions from both the true RV dispersion of the 
cluster and undetected binaries, which also artificially inflates the observed RV dispersion.
We aim to recover the true RV dispersion from our data.

First we calculate the observed RV dispersion, which is simply the standard deviation of the observed mean RVs for each 
star about the cluster mean RV (calculated in Section~\ref{sub:membership}).
In our calculation, we use the weighted mean RVs for each star in the given sample (see Table~\ref{RVtab}), which utilize 
the single-measurement precision values from the given telescope for each individual RV value (see Section~\ref{sub:prec}).
For the single main-sequence, subgiant and giant stars in our primary sample, the resulting observed velocity dispersion is $0.86 \pm 0.11$~\kms.
After correcting for the observational error (following the method of \citealt{mcnamara:77} and \citealt{mcnamara:86b}), we find a combined RV dispersion 
of $0.80 \pm 0.04$~\kms.

We then follow a similar method to that of \citet{geller:10} to correct this combined RV dispersion for the contribution from 
undetected binaries.  Briefly, we run a Monte Carlo analysis to create many realizations of our M67 observations.
For each realization, we generate a population of synthetic single and binary stars (given an input binary frequency), and 
produce synthetic RVs for these stars on the true observing dates from our M67 survey, which we analyze in the same manner as our real observations.  
We run this analysis for a range of true RV dispersion values, and for each value, we derive the difference between the
synthetic combined RV dispersion and the input true RV dispersion.  This difference is the contribution from undetected binaries
\citep[$\beta$ in Equation~5 of][]{geller:10}.

For this analysis, we improve upon the technique of \citet{geller:10} in our treatment of the binaries, and therefore, for clarity, we  explain the method in some further detail here.
In order to recover the true velocity dispersion, we must also estimate the binary frequency of the cluster.  
We use a similar Monte Carlo method to \citet{geller:12} to
account for our incompleteness in binary detections out to the hard-soft boundary.  
As in \citet{geller:12}, we assume that the orbital parameters of the M67 binaries follow the same distributions 
as observed for the solar-type binaries in the Galactic field from \citet{raghavan:10}, except 
here we use a circularization period of 12.1 days \citep{meibom:05}.
We also make a few updates to the method of \citet{geller:12}.  First, rather than choosing one specific primary mass, we draw from an inferred primary-mass distribution
within our M67 primary sample (derived by comparisons to a \citealt{marigo:08} isochrone).
Second, we attempt to recreate the correlation between binary mass ratio and orbital period observed for field binaries by 
\citet[][see Figure~17 and related discussion]{raghavan:10}.  Specifically, for binaries with periods between 100 days and 1000 years,
we choose mass ratios from a uniform distribution with 10\% of systems having a mass ratio of unity (e.g., twins).  
For binaries with periods greater than 1000 years we draw mass ratios from a uniform distribution limited to be less than 0.95.  For 
binaries with periods less than 100 days, we enforce $\sim$19\% to have mass ratios between 0.2 and 0.45, $\sim$44\% to have mass ratios between 0.45 and 0.9, 
and the remainder to have mass ratios between 0.9 and 1, all drawn from uniform distributions between the respective mass ratio limits.
Finally, we limit the field log-normal orbital period distribution so that the binaries are detached (using radii estimates from a \citealt{marigo:08} isochrone), and 
the periods are less than the hard-soft boundary.  

The assumed location of the hard-soft boundary, which is at roughly $10^5$-$10^6$ days in M67, is especially important for the undetected binary correction. 
We do not expect to detect binaries at these long periods.  Furthermore, the hard-soft boundary is near the peak of the log-normal distribution, and therefore
the assumed binary frequency is particularly sensitive to this cutoff.  
We estimate the hard-soft boundary as the location where a synthetic binary's binding energy is equal to the kinetic energy of a ``typical'' star moving at an assumed
velocity dispersion.  For the mass of this typical star, we take the mean mass of an object (single or total mass of a binary) 
from the \citet{hurley:05} $N$-body simulation of M67, which at 4 Gyr is 0.95~\Msolar.  
We then assume a velocity dispersion, run our analysis to derive the true velocity 
dispersion in our M67 sample (given the resulting hard-soft boundary and total binary frequency), and iterate until this assumed velocity dispersion is $<0.05$~\kms\ different 
from the derived true velocity dispersion in M67 (a somewhat arbitrary limit meant to be roughly equivalent to the precision with which we can measure 
the velocity dispersion).  As a starting guess, we use the combined dispersion value of 0.80~\kms, found above.  

Our analysis of the single main-sequence, giant and sub-subgiant stars in this sample requires only two iterations.  
The resulting orbital period distribution matches closely to those predicted by the $N$-body open cluster models of \citet{hurley:05} and \citet{geller:13}, and
we estimate the total binary frequency in our sample to be 57\%~$\pm$~4\%.  We derive a true velocity dispersion, after correcting for 
measurement error and undetected binaries, for the M67 single main-sequence stars, subgiants and giants with $V \leq 15.5$ of $0.59 \substack{+0.07\\-0.06}$~\kms.

Using a subsample of 20 of the brightest stars in our sample, \citet{mathieu:83} measured a cluster RV dispersion, after correcting for measurement errors,
of $0.48 \pm 0.09$~\kms.  After also correcting for undetected binaries (assuming a 50\% binary frequency), \citet{mathieu:83} find an 
RV dispersion of $0.25 \pm 0.18$~\kms.
\citet{girard:89} show that this RV dispersion increases to $0.48 \pm 0.15$~\kms\ (corrected for both measurement errors and undetected binareis) when including 
the larger sample from \citet{mathieu:86}, in good agreement with the result we find here.  

More recently, \citet{pasquini:12} derive an RV dispersion for main-sequence stars in M67 with $V<15$ of $0.680 \pm 0.063$~\kms, and for M67 giants of $0.540 \pm 0.090$~\kms,  
using HARPS spectra with a typical precision of $\sim$10 m s$^{-1}$\ (not corrected for undetected binaries).
If we divide our sample into the SM main-sequence stars with $V<15$ and SM giant stars,
after correcting for measurement errors, we find a combined velocity dispersion for the main-sequence stars
of $0.83 \pm 0.04$~\kms\ and for giants of $0.60 \pm 0.08$~\kms.
However, the main-sequence stars have a much higher binary frequency than the giants, and therefore a much larger correction for undetected binaries.
After correcting for both measurement errors and undetected binaries, 
we find a true velocity dispersion for main-sequence stars ($V<15$) 
of $0.60 \substack{+0.08\\-0.07}$~\kms\ and for giants of $0.60 \pm 0.10$~\kms.  Thus we find essentially identical velocity dispersions for the main-sequence and giant stars.
We note that Padova isochrones indicate a mean mass for the main-sequence stars in this sample of 1.11~\Msolar\ (with a standard deviation of 0.10 \Msolar), and for
the giants of 1.33~\Msolar\ (with a standard deviation of 0.01 \Msolar); our sample covers only a narrow mass range.

One-dimensional proper-motion dispersion measurements
from the literature are somewhat higher than these RV dispersions \citep[as was also noted by][]{girard:89}.  
\citet{mcnamara:78} find a velocity dispersion of 0.95~\kms\ with a $1 \sigma$ upper limit of $<$1.48~\kms. %, at a distance of 843 pc.
\citet{zhao:96} find a dispersion of $0.96 \pm 0.09$~\kms, and
\citet{girard:89} find a dispersion of $0.81 \pm 0.10$~\kms.
As discussed above, the \citet{girard:89} proper motions have the highest precision, and therefore here we will compare directly to their result.
After correcting for undetected binaries, our true RV dispersion measurement differs from the \citet{girard:89} proper-motion dispersion measurement at about the $2 \sigma$ level
(without accounting for an additional uncertainty on the proper-motion dispersion from the range in cluster distances quoted in the literature).
Given the uncertainties on these measurements, the different distances assumed in converting from angular proper motions to \kms, and the different methods 
used to account for measurement error in the proper-motion dispersions, we conclude that the RV and proper-motion dispersions are in agreement at our level of precision.

\begin{figure}[!t]
%\epsscale{0.6}
\plotone{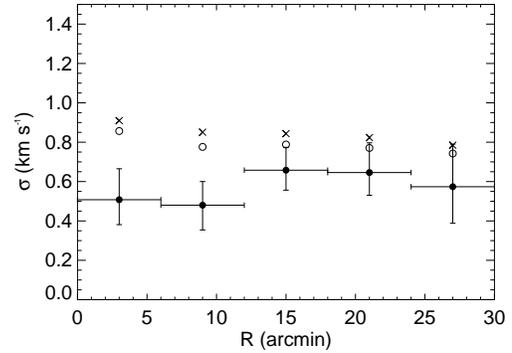}
\epsscale{1.0}
\caption{\small Radial-velocity dispersion as a function of radius from the 
cluster center.  Only single main-sequence stars, subgiants and giants with $V \leq 15.5$ are used for this analysis. 
The observed radial-velocity dispersions are shown with crosses,
combined radial-velocity dispersions (after correcting for the contribution from measurement error)
are shown in open circles, and the true radial-velocity 
dispersions (after also correcting for undetected binaries) are shown in filled circles.
The horizontal bars show the range in radius of each bin, and the vertical bars
show the uncertainties on the velocity dispersion values. All uncertainties are 
derived as in \citet{geller:10}.}
\label{fig:vdisp}
\end{figure}

We have also examined the true RV dispersion of M67 as a function of radius from the cluster center, shown in Figure~\ref{fig:vdisp}.
We divide this sample of main-sequence, subgiant and giant single members into five equal bins in radius.
Crosses, open circles and filled circles show the observed, combined and true RV dispersion values, respectively, at each bin in 
radius.  The correction for undetected binaries is highest in the inner-most bin, as this bin has the highest binary frequency,
consistent with our finding that the binaries are mass segregated with respect to the single stars (see Figure~\ref{fig:spatdist}).

The radial distribution of the true velocity dispersion 
within the parameter space covered by our survey is consistent with an isothermal distribution.
\citet{zhao:96} find an increase in the proper-motion dispersion as a function of radius,  
although they note that this effect may be in part due to increased field-star contamination with radius in their sample.
We do not see a similar trend in our data.  We have also examined different binnings with no 
detectable effect on the results presented here.

Given the velocity dispersion, we can estimate the mass of the cluster through the virial theorem.  Under the assumption that the cluster is 
in dynamical equilibrium, the kinetic and potential energies are related by $V^2 = \eta G M / R$.  From \citep{spitzer:87}, if this 
equation is rewritten in terms of the half-mass radius ($r_{\rm h}$), $\eta \sim 0.4$ for most systems.  Also, for an isotropic velocity dispersion, 
the observed half-mass radius in projection is given by $r_{\rm hp} \sim \frac{3}{4}r_{\rm h}$, and the one-dimensional RV dispersion is 
related to the rms velocity $V$ by $\sigma_r^2 = V^2/3$.  Thus the virial mass of the cluster can be calculated by the equation:

\begin{equation}
M_V = 10 \frac{r_{\rm hp} \sigma_r^2}{G} .
\end{equation}

\citet{fan:96} find a half-mass radius of M67 for stars with $13.8 < V < 14.5$ of 
$2.54 \pm 0.41$~pc
and for stars with $15.6 < V < 14.5$ of 
$2.74 \pm 0.29$~pc,
assuming a distance of 850 pc. The M67 sample used here extends from $V \sim 8$ to $V = 15.5$, though the vast majority of the stars have $V \gtrsim 12.5$.  
We will assume a projected half-mass radius for the stars in our sample of 
$r_{\rm hp} = 2.58 \pm 0.45 $~pc.
This value lies in the middle and extends to the limits of the range in $r_{\rm hp}$ found by \citet{fan:96} for these two magnitude regimes.
Given our true RV dispersion above of $\sigma_r = 0.59 \substack{+0.07\\-0.06}$~\kms, we find a virial mass for M67 of 
$2100 \substack{+610\\-550}$~\Msolar.

This value is in good agreement with previous dynamical mass estimates.
\citet{mcnamara:78} derive a virial mass of 1600~\Msolar\ (with a $1 \sigma$ upper limit of 4000~\Msolar), 
and a direct cluster mass of 1100~\Msolar\ based on star counts.  
\citet{zhao:96} find a virial mass of $1500 \pm 250$~\Msolar\ using their proper motions.
From examination of the cluster luminosity function, \citet{fan:96} calculate a mass of 1270~\Msolar\ within a radius of 2000 arcsec and including stars 
with masses $>0.5$~\Msolar.

These mass estimates (including ours) are somewhat higher than the mass estimates of \citet{montgomery:93}, \citet{francic:89}
and \citet{mathieu:83}, who find cluster masses of 724~\Msolar, 553~\Msolar\ and 903~\Msolar, respectively, all through analyses of the cluster
luminosity function.  We interpret these measurements as lower limits 
on the true cluster mass, as each of these surveys were limited either in magnitude or radial extent.

\section{Summary}
\label{sec:summary}

This is the first in a series of papers studying the dynamical state of the old open cluster M67 through
precise RVs.  Here we focus on determining cluster membership and identifying binaries (Sections~\ref{sec:results}) within a 
complete sample of cluster stars.
In total, we present results from \NRV\ RV measurements of \Nstars\ stars in the direction of M67 spanning
from the brightest stars in the cluster down to about 4 magnitudes below the main-sequence turnoff ($V = 16.5$), 
covering a mass range of about 1.34~\Msolar\ to 0.76~\Msolar, and extending 
spatially to 30 arcmin in radius from the cluster center (about 6 or 7 core radii).  We combine RV measurements from multiple telescopes using different
instruments, carefully accounting for their different single-measurement RV precisions (see Section~\ref{sub:prec}).  The vast majority 
of the stars in our sample have multiple epochs of RV measurements, allowing for the detection of binary companions.  For some stars,
the time span of our observations exceeds 40 years, and about one third of these stars have observations spanning at least 10 years.

In our analysis, we require at least 3 RV measurements before attempting to determine 
RV variability or RV membership probabilities.  We also utilize the proper-motion membership surveys of \citet{sanders:77}, \citet{girard:89}, \citet{zhao:93} and \citet{yadav:08}
to remove additional non-members from our cluster sample.
Using both proper-motion and RV membership information, we identify \NMEM\ cluster members within our sample, \NVAR\ of which show 
significant RV variability, indicative of a binary companion (or multiple companions).  

We define a primary sample of stars with $V \leq 15.5$ and within 30 arcmin from the cluster center, where we have $\geq$3 RV measurements
for all proper-motion members 
(and, considering the entire primary sample, at least one RV measurement for all but 1 star and $\geq$3 RV measurements for all but 4 stars; see Section~\ref{sub:comp}).  
We then use this primary sample to construct a CMD cleaned from field-star
contamination (Section~\ref{sub:CMD}), identify and discuss a few notable stars (Section~\ref{sub:snote}),
compare the projected radial distributions of different cluster populations 
(Section~\ref{sub:spatdist}), and determine the RV dispersion and virial mass of the cluster (Section~\ref{sub:vdisp}).

Within our cluster member sample, we identify 14 BSS, 4 bright ``yellow giants''
residing on the CMD between the typical BSS region and the giant sequence  (though one is outside of a 30 arcminute radius from the cluster center), 
and 2 ``sub-subgiants'' located to the red of the main-sequence but fainter than the subgiant and giant branches.  
These ``exotic'' stars have a remarkably high binary frequency of (at least) 80\%, as compared to 22.7\% detected binaries amongst the rest of the sample.

Both the binaries and BSS are significantly more centrally concentrated than the single stars (i.e., non-RV variables)
in our sample.  Within our sample the binaries are, on average, more massive than the single stars, and therefore this result 
confirms that M67 is mass segregated (as also found by other authors), which is expected for a cluster that 
has lived through tens of relaxation times.  As \citet{mathieu:86b} also discuss, the central concentration of the BSS
suggests that they too are more massive than the single stars in our sample, which is consistent with the predictions of 
theoretical formation channels.  We do not observe a bimodal radial distribution for the BSS,
as is observed for BSS in many globular clusters and those in the old open cluster NGC 188.
However, our sample only extends to roughly 6 or 7 core radii,
and a larger radial extent would likely be required to detect a possible halo BSS population and bimodal radial distribution.

Finally, we determine the RV dispersion of the single main-sequence, subgiant and giant members in our primary sample.  
Accounting for measurement errors, we find a
combined RV dispersion of $0.80 \pm 0.04$~\kms.  When also corrected for the contribution from undetected binaries, 
we find a true RV dispersion of the cluster of $0.59 \substack{+0.07\\-0.06}$~\kms.  The radial distribution of the true RV dispersion within our sample
is consistent with an isothermal distribution.  Using this true RV dispersion and a projected half-mass radius of $r_{\rm hp} = 2.58 \pm 0.45 $~pc, we calculate
a virial mass for M67 of $2100 \substack{+610\\-550}$~\Msolar.

Our long-term RV survey of M67 enables a detailed study of the cluster's binary population, from the main-sequence through 
the giant branch and including the rich population of M67 BSS.
Future papers in this series will focus on the binary properties of the cluster (e.g., binary frequency and distributions 
of orbital elements).  In particular, we have determined orbital solutions for \BM\ binary members of the cluster, which we 
will present and analyze in subsequent papers.  Additionally, we will study the binary properties of the BSS in detail, 
which are critical for our understanding of their formation mechanism(s) \citep[e.g.][]{geller:11, mathieu:15}.  
Indeed, binary stars play a primary role in the dynamical evolution of star clusters and the formation of exotic stars like BSS.  
With the addition of the binary properties to the large body of existing observational work on the cluster, M67 will 
be invaluable to our theoretical understanding of stellar dynamics, stellar evolution, the formation of BSS, and the 
long-term evolution of star clusters.

\acknowledgments
The authors would like to thank the many individuals who helped obtain these spectra and determine the stellar 
radial velocities, both at the CfA:
Jim Peters,
Bob Davis,
Ed Horine,
Perry Berlind,
Ale Milone,
Robert Stefanik,
Mike Calkins,
John Geary,
Andy Szentgyorgyi,
Gabor Furesz,
and at the University of Wisconsin - Madison:
Natalie Gosnell,
Katelyn Milliman,
Emily Leiner,
Ben Tofflemire.
We would also like to express our gratitude to the staff of the WIYN 
Observatory for their skillful and dedicated work that have allowed us to 
obtain these excellent spectra.   We thank Imants Platais for his help in cross referencing between
the various catalogs in the literature for M67.
We also thank any other undergraduate and 
graduate students not mentioned here explicitly who have contributed late nights to obtain the spectra for 
this project.  A.M.G. is funded by a National Science Foundation Astronomy 
and Astrophysics Postdoctoral Fellowship under Award No. AST-1302765.
This work was also supported by NSF grant AST 0406615 and the 
Wisconsin Space Grant Consortium.

Facilities: \facility{Tilinghast 1.5m, WIYN 3.5m, MMT, Palomar Hale 5m, Wyeth 1.5m, Haute Provence}

\bibliographystyle{apj}

%\tabletypesize{\tiny}
\clearpage
\begin{landscape}
\setlength{\tabcolsep}{2pt}
\begin{deluxetable}{lllccccccccccccccccccll}
\tablecaption{\label{RVtab}Radial-Velocity Data Table}
\tablehead{&&&&&&&&&&&&&&&&&&&&&&\\[4pt]
\colhead{$ID_{\rm W}$} & \colhead{$ID_{\rm X}$} & \colhead{$RA$} & \colhead{$Dec$} & \colhead{$V$} & \colhead{$(\bv)$} & \colhead{$N_{\rm W}$} & \colhead{$N_{\rm C}$} & \colhead{$JD_0$} & \colhead{$JD_f$} & \colhead{$\overline{RV}$} & \colhead{$RV_e$} & \colhead{$i$} & \colhead{$P_{\rm RV}$} & \colhead{$P_{\rm PMy}$} & \colhead{$P_{\rm PMz}$} & \colhead{$P_{\rm PMg}$} & \colhead{$P_{\rm PMs}$} & \colhead{$e/i$} &  \colhead{$P(\chi^2)$} & \colhead{Class} & \colhead{Comment}}\\[1pt]
%\rotate
%\tablewidth{0pt}
\startdata
&&&&&&&&&&&&&&&&&&&&&&\\[4pt]

  1001 &   S1024 &  8:51:22.91 & 11:48:49.4 & 12.720 &   0.550 &   0 &  34 & 45784.84 & 47519.82 &   33.32 &    0.17 & \nodata &      98 &      99 & \nodata &      99 &      92 & \nodata & \nodata &      BM & SB2,CX111,PV \\[1pt]
  2001 &   S1027 &  8:51:24.95 & 11:49:00.8 & 13.240 &   0.600 &  14 &  13 & 46808.93 & 56707.92 &   32.09 &    0.15 &    0.67 &      93 &     100 &      93 &      99 &      95 &    1.18 &   0.000 &      SM & X46 \\[1pt]
  3001 &   S1031 &  8:51:22.96 & 11:49:13.1 & 13.260 &   0.460 &   2 &  32 & 46808.96 & 54423.99 &   33.46 &    0.30 &    0.91 &      98 &     100 &      94 &      99 &      91 &    7.76 &   0.000 &      BM & SB1 \\[1pt]
  4001 &   S1029 &  8:51:21.62 & 11:49:02.5 & 15.210 &   0.790 &  17 &   7 & 47580.74 & 56705.91 &   32.60 &    0.26 &    0.79 &      97 &      97 & \nodata &      72 &      83 &   11.62 &   0.000 &      BM & SB1,PV \\[1pt]
  5001 &   M5754 &  8:51:23.50 & 11:49:05.8 & 16.220 &   1.000 &   3 &   0 & 56054.68 & 56669.01 &   34.24 &    0.27 &    0.74 &      98 &      96 & \nodata & \nodata & \nodata &    0.63 &   0.669 &      SM & \nodata \\[1pt]
  1002 &   S1023 &  8:51:26.84 & 11:48:40.5 & 10.540 &   0.570 &   0 &   7 & 41072.69 & 50822.92 &    3.92 &    0.38 &    0.49 &       0 &      84 &       0 &       7 &       0 &    2.04 &   0.000 &      SN & \nodata \\[1pt]
  2002 &   S1040 &  8:51:23.77 & 11:49:49.3 & 11.520 &   0.870 &   1 &  57 & 41073.67 & 54164.75 &   33.01 &    0.08 &    0.56 &      98 &      97 & \nodata &      99 &      95 &   10.05 &   0.000 &      BM & SB1,X10,CX6,PV,YG \\[1pt]
  3002 &   S1018 &  8:51:24.09 & 11:48:21.9 & 12.830 &   0.570 &   0 &  32 & 46874.69 & 48291.90 &   33.38 &    0.16 & \nodata &      98 &     100 &      96 &      99 &      89 & \nodata & \nodata &      BM & SB2 \\[1pt]
  4002 &   S1030 &  8:51:25.95 & 11:49:08.9 & 13.230 &   0.570 &   2 &   5 & 46808.94 & 56051.71 &   34.22 &    0.22 &    0.77 &      98 &     100 & \nodata &      98 &      90 &    0.76 &   0.653 &      SM & \nodata \\[1pt]
  5002 &   S1032 &  8:51:26.52 & 11:49:20.3 & 13.480 &   0.570 &   2 &   4 & 47489.97 & 55934.05 &   34.24 &    0.35 &    0.75 &      98 &     100 & \nodata &      98 &      96 &    1.15 &   0.169 &      SM & PV \\[10pt]
\enddata

\end{deluxetable}
The contents of each column are defined in Section 5.

\setlength{\tabcolsep}{6pt}
\clearpage
\end{landscape}

\end{document}